\documentclass[preprint,review,12pt]{elsarticle}

\usepackage{amssymb}
\usepackage{amsmath}

\usepackage{tabularx}
\usepackage{xcolor}
\usepackage{url}
\usepackage{booktabs}
\usepackage{graphicx} 
\usepackage[most]{tcolorbox}
\usepackage{lipsum}
\usepackage{tabularx}
\usepackage{booktabs}
\usepackage{geometry}
\usepackage{longtable}
\usepackage{array}
\usepackage{caption}
\usepackage{graphicx}
\usepackage{subcaption}
\newtcolorbox{paperexamplebox}[1]{%
  enhanced,
  breakable,
  colback=gray!3,
  colframe=gray!55!black,
  boxrule=0.6pt,
  arc=2mm,
  left=6pt,right=6pt,top=6pt,bottom=6pt,
  title=\textbf{#1},
  fonttitle=\bfseries,
}

\usepackage{float}
\usepackage{adjustbox}   

\journal{Nuclear Physics B}

\begin{document}

\begin{frontmatter}



\title{Empathic Prompting: Non-Verbal Context Integration for Multimodal LLM Conversations} 





\author[label1]{Lorenzo Stacchio$^*$}
\author[label2]{Andrea Ubaldi}
\author[label3]{Alessandro Galdelli}
\author[label2]{Maurizio Mauri}
\author[label1]{Emanuele Frontoni}
\author[label2]{Andrea Gaggioli}

\affiliation[label1]{organization={University of Macerata},
            city={Macerata},
            country={Italy}}

\affiliation[label2]{organization={Università Cattolica del Sacro Cuore},
            city={Milan},
            country={Italy}}

\affiliation[label3]{organization={Università Politecnica delle Marche},
            city={Ancona},
            country={Italy}}

\cortext[cor1]{Corresponding author: lorenzo.stacchio@unimc.it}

\begin{abstract}
We present Empathic Prompting, a novel framework for multimodal human–AI interaction that enriches Large Language Model (LLM) conversations with implicit non-verbal context. The system integrates a commercial facial expression recognition service to capture users’ emotional cues and embeds them as contextual signals during prompting. Unlike traditional multimodal interfaces, empathic prompting requires no explicit user control: instead, it unobtrusively augments textual input with affective information for conversational and smoothness alignment. The architecture is modular and scalable, allowing integration of additional non-verbal modules. We describe the system design, implemented through a locally deployed DeepSeek instance, and report a preliminary service and usability evaluation (N=5). Results show consistent integration of non-verbal input into coherent LLM outputs and maintaining a coherent and contextually appropriate exchange. Beyond this proof of concept, empathic prompting points to applications in chatbot-mediated communication, particularly in domains like healthcare or education, where users’ emotional signals are critical yet often opaque in verbal exchanges.
\end{abstract}

\begin{graphicalabstract}
\includegraphics[width = \linewidth]{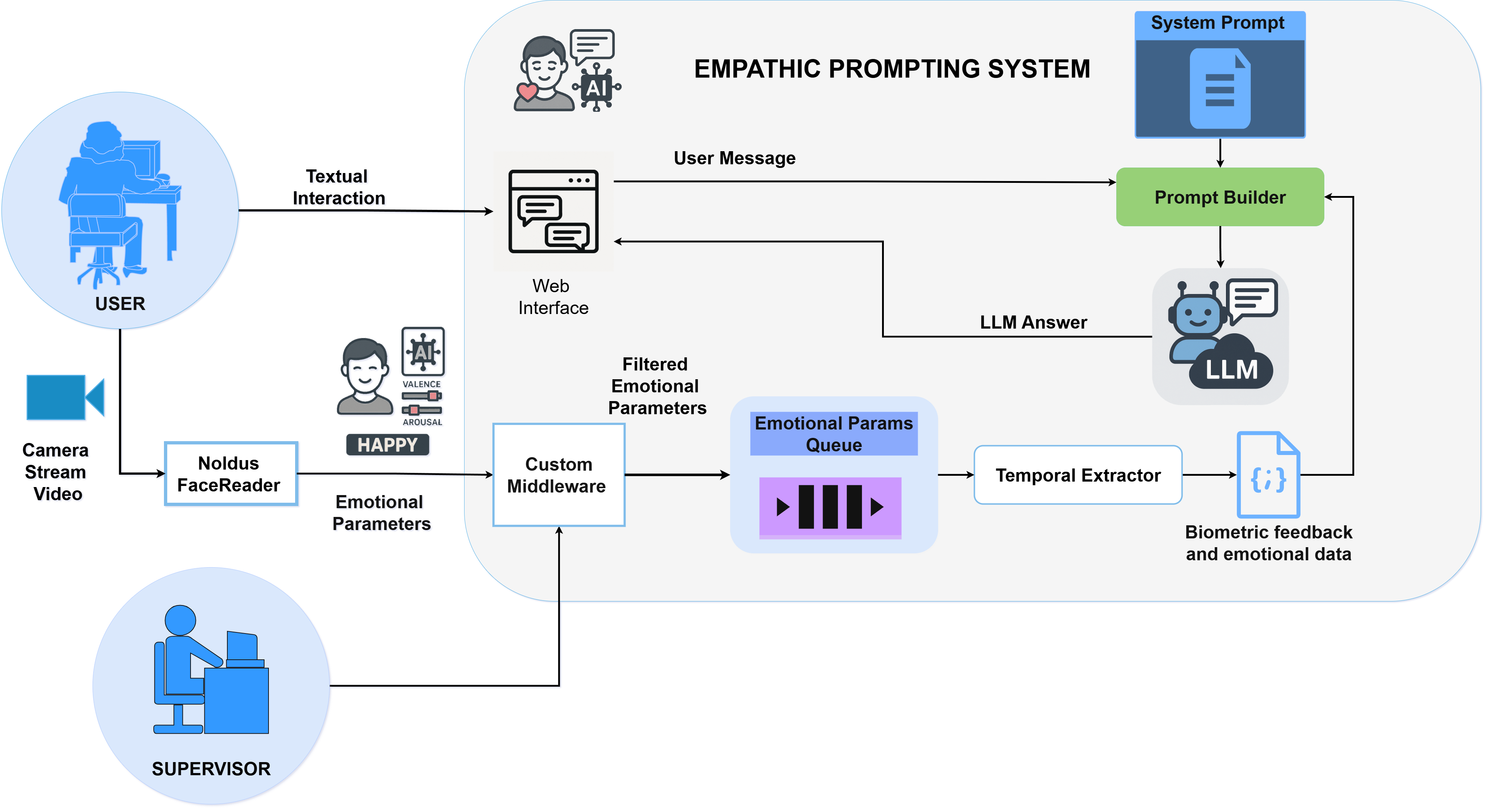}
\end{graphicalabstract}

\begin{highlights}
\item Empathic Prompting fuses LLMs with non-verbal cues for adaptive responses.
\item A multimodal pipeline integrates face-detected emotions with textual signals in real-time.
\item A novel LLM-based evaluation protocol measuring emotional conversations.
\item A pilot study with human experts to measure empathy and naturalness.
\end{highlights}

\begin{keyword}
Empathic Artificial Intelligence\sep Non-verbal communication \sep Large Language Models\sep Human-in-the-loop \sep Affective computing
\end{keyword}

\end{frontmatter}



\section{Introduction}

Empathy is increasingly recognized as a central ingredient of human–Artificial Intelligence (human-AI) interaction \cite{inzlicht2024praise}, especially in domains such as health, education, and psychological wellbeing, where trust, rapport, and engagement are essential outcomes. In human communication, empathy emerges through both cognitive appraisals and affective resonance, manifesting through a continuous interplay of verbal and nonverbal cues \cite{Cooper2016Empathy}. Classic and contemporary evidence demonstrates that nonverbal channels (e.g., eye contact, posture, body orientation, interpersonal distance) substantially contribute to judgments of empathy, often outweighing verbal content alone and underscoring the embodied foundations of empathic communication \cite{haase1972nonverbal}.

Concurrently, large language models (LLMs) have demonstrated remarkable capabilities in generating responses that humans perceive as empathic \cite{ayers2023comparing, inzlicht2024praise}. Recent systematic reviews indicate that LLMs exhibit elements of cognitive empathy, including emotion recognition and supportive language generation, with some responses preferred over human responses in medical contexts, though significant methodological and ethical evaluation challenges remain \cite{sorin2024large}. These converging developments motivate a multi-modal perspective on empathy in human–AI interaction: if empathic communication in humans fundamentally depends on coordinated verbal and nonverbal channels, then conversational AI systems should be designed to integrate nonverbal signals directly into the language generation process \cite{arjmand2024empathic}.

\textcolor{black}{Current research at the intersection of affective computing and generative models has made significant advances in this direction \cite{Li2025EmoVerse,Liu2024OpenSet,Han2024Knowledge}, suggesting that models can be engineered to classify, reason about, and contextualize emotions effectively~\cite{dongre2024physiology,neupane2025wearable}.} However, these approaches primarily focus on recognition and evaluation benchmarks, while achieving real-time conversational alignment with human interlocutors is a less developed area. More closely aligned with interaction design concerns, recent work by \cite{arjmand2024empathic} on "Empathic Grounding" explores a computational model where empathy functions as grounding evidence within embodied dialogue systems, though their approach targets robotic behavior planning rather than core language interfaces.

Building on this emerging foundation, our contribution shifts the integration locus from behavior planning to prompting itself as the primary mechanism through which nonverbal context shapes empathic language generation. 
We term this approach \textit{Empathic Prompting}~\footnote{A version of this work is available as a preprint at \cite{Anonymous2025EmpathicPrompting}}.
Our framework incorporates nonverbal signals—estimates derived from facial expressions capturing valence, arousal, and basic emotions—directly into the prompt pipeline of conversational LLMs. Crucially, model parameters remain unchanged; instead, the system conditions are generated at inference time by embedding structured affective context within the system and message prompts. This design is aimed at transforming implicit nonverbal communication into explicit semantic representations that can modulate conversational tone, supportive strategy selection, and emotional alignment without requiring specialized training data or architectural modifications, leveraging the well-documented sensitivity of LLMs to prompt structure and content \cite{li2023large}.

We hypothesize that this strategy could confer significant advantages in LLM applications within sensitive domains, such as medicine, mental health, and education, where empathic alignment is critical to user acceptance and therapeutic impact. \textcolor{black}{In these contexts, enriching the language interface with nonverbal information about users' affective states can improve both conversational fit and safety, enabling more contextually appropriate responses that align with users' emotional needs (i.e., improving conversational fluidity~\footnote{\textcolor{black}{Conversational fluidity denotes dialogic coherence and interactional flow and not real-time latency.}}~\cite{skantze2021turn})}.

The Empathic Prompting framework operates through three integrated functions: (i) sensing, which extracts affective descriptors from facial expressions; (ii) mapping, which converts these signals into transparent semantic descriptors combining valence and arousal ranges with canonical emotion terms; and (iii) prompt enrichment, which integrates these descriptors into system prompts and message histories using lightweight control heuristics for safety and tone modulation. 

This architecture offers three key advantages: modularity and agnosticism to particular LLMs or sensing toolkits enable rapid cross-platform portability; the semantic nature of affective descriptors facilitates auditing and human-in-the-loop oversight; compatibility with ongoing advances in affect recognition \cite{Liu2024OpenSet}, multitask multimodal emotion understanding \cite{Li2025EmoVerse}, and knowledge-based context integration \cite{Han2024Knowledge}, which allows the framework to evolve with the field.
To evaluate this approach, we pose two research questions:
\begin{itemize}
\item \textbf{RQ1}: Does integrating non-verbal context through prompting improve perceived empathy and safety-aligned behavior? 
\item \textbf{RQ2}: How do users experience conversational smoothness and alignment when generation is conditioned on real-time affective signals? 
\end{itemize}

We address these questions through a mixed-methods evaluation combining an internal usability study (N=5) with an LLM-as-a-Judge protocol organized around three rubrics (i.e., Empathy Support, Safety Boundary, and System-Prompt Adherence) to ensure explicit and auditable evaluation. Preliminary findings suggest that non-verbal context is consistently integrated into responses and effectively shapes both tone and supportive strategies, though we also identify failure modes, including verbosity and occasional misaligned suggestions that highlight important design trade-offs.

\textcolor{black}{This paper proceeds as follows: Section \ref{sec:background} reviews the theoretical foundations and empirical advances motivating our approach, then presents the Empathic Prompting framework and implementation details (Section ~\ref{sec:system}), followed by our evaluation methodology and findings (Sections ~\ref{sec:use_case} and ~\ref{sec:results}, respectively). Finally, Section \ref{sec:disc_concl} concludes the work, discussing the obtained results and design implications, ethical considerations, and recommendations for future large-scale, IRB-approved studies centered on user experience and outcome measures.
}

\section{Background}
\label{sec:background}
Empathy means attuning to another person’s feelings, adopting their perspective, and fostering a sense of compassionate care \cite{decety2014complex}. Contemporary analyses of empathy converge on four interrelated themes that together capture its conceptual core. First, empathy entails \textit{understanding} another person’s inner world, a cognitive dimension that involves perspective-taking and recognizing the other’s mental and emotional states. Second, empathy involves \textit{feeling}, that is, an affective resonance with the other’s experiences, whereby one becomes emotionally attuned to the situation of the other. Third, empathy includes \textit{sharing}, namely the capacity to experience states similar to those of the other person, such that the empathizer enters into the other’s experiential world “as if” it were their own. Finally, empathy requires \textit{self–other differentiation}, a recognition that the source of the feelings or experiences belongs to the other and not to oneself \cite{eklund2021toward}. Crucially, empathy operates as an interpersonal phenomenon that unfolds through dynamic interaction between empathizer and target, involving understanding, emotional recognition, perceived similarity, and concern expressed through concrete actions \cite{haakansson2003empathy}, rather than as a static trait residing solely within the empathizing individual. 

Early research in therapeutic and laboratory settings suggested that nonverbal cues—such as prosody, timing, facial expression, gaze, posture, and micro backchannels—can strongly shape perceptions of empathic communication, although these findings were derived from highly constrained experimental contexts \cite{haase1972nonverbal,mehrabian1967decoding}. Subsequent studies across diverse settings have consistently highlighted the importance of nonverbal channels in conveying affective and relational information \cite{hall1995nonverbal,bavelas2000listeners}, while research on interpersonal synchrony shows that temporal alignment in movement and prosody fosters perceived understanding and rapport, with disruptions in synchrony potentially undermining empathic attunement \cite{tickle1990nature,chartrand2013antecedents}.

Understanding empathy as a multimodal process sets the stage for evaluating how, and to what extent, artificial systems such as Large Language Models (LLMs) might approximate or support empathic communication. LLMs are autoregressive neural networks trained at scale to predict the next token in context; instruction tuning and reinforcement learning from human feedback shape them into general-purpose assistants capable of following natural language directives and producing contextually appropriate text. With vision–language extensions, some LLMs can also consume images (and, in emerging systems, audio), enabling multimodal reasoning. Because they can parse intent, maintain discourse coherence, and modulate style, LLMs are increasingly explored in medicine, mental health, psychoeducation, and learning-support settings \citep{ayers2023comparing, abd-alrazaq2020effectiveness, kasneci2023chatgpt, kuhail2023interacting}. In these domains, access to emotional information is not ornamental: it underpins trust, adherence, de-escalation, and engagement; guides triage and signposting; calibrates tone and directiveness; and supports tailoring (e.g., pacing, question framing, and the balance of validation vs. action-oriented guidance).
Text-only empathic inference is, however, intrinsically limited. Emotional states are often implicit; users may struggle to label them (e.g., alexithymia) or deliberately downplay them in writing. Prosodic and kinesic cues, central to empathic appraisal, are absent in plain text. Lexical markers of affect are noisy and culturally variable; irony, sarcasm, and politeness strategies confound simple sentiment heuristics. Emojis and lexical valence–arousal–dominance (VAD) cues help but are ambiguous and context-dependent, and numeric affect scales injected into prompts can misalign with users’ lived intensity. \textcolor{black}{These limitations motivate designs that integrate nonverbal context into the conversational loop rather than infer emotion from text alone~\cite{dongre2024physiology,neupane2025wearable}.}

A growing body of HCI research demonstrates the promise of multimodal approaches to fostering empathic interaction. A particularly relevant example is offered by Arjmand et al. \cite{arjmand2024empathic}, who introduce the concept of “empathic grounding” as an extension of Clark's \cite{clark1991grounding} theory of conversational grounding, in which the listener's empathy for the speaker's affective state becomes part of the grounding criterion. Their computational model processes two streams of user input: verbal utterances obtained through speech transcripts, and facial expressions summarized into the most salient emotion labels. These inputs are passed to a LLM, which then generates multimodal grounding moves that combine verbal responses, affective states, and nonverbal behaviors such as head movements and facial displays. In a controlled testbed using a humanoid robot to interview participants about past pain episodes, their between-subjects experiment (N=18) found that participants in the empathic grounding condition—compared to a baseline condition using neutral backchannels—rated the agent significantly higher on empathy, emotional intelligence, trust, working alliance, and perceived understanding. While the study was conducted in a highly constrained domain with Wizard-of-Oz control, the findings suggest that empathic alignment may benefit from explicitly modeling and communicating affect across modalities, establishing empathic grounding as a promising framework for integrating LLMs with multimodal sensing in empathic AI design.

Our work builds on these insights but shifts the focus of integration from behavior planning to the language interface itself. We propose to treat nonverbal affective signals (e.g., valence, arousal, basic emotions) as structured context injected directly into the prompt, so that language generation is conditioned on real-time emotional cues without requiring model retraining. This Empathic Prompting perspective regards nonverbal signals as first-class conversational context, enabling the recovery of key ingredients of empathic interaction—tone calibration, pacing and turn-taking, supportive strategy selection, and the nuanced handling of verbal–nonverbal incongruence—within a transparent and auditable prompting pipeline. Importantly, we do not claim that LLMs can possess genuine empathic capacities; consistent with Inzlicht and colleagues \cite{inzlicht2024praise}, our stance is pragmatic. Consistently, the design question we pursue is instrumental: how can an LLM harness users’ emotional signals to convey an empathic tone? More broadly, how can it adapt its responses to the user’s state by treating emotional information as extra-linguistic context for generation, while remaining transparent about its fundamentally non-human status?

We address these challenges through the following contributions. First, we enable real-time integration of non-verbal affective signals into prompt conditioning, allowing emotional context to directly influence language generation at inference time. Second, our semantic mapping approach converts biometric data into transparent, interpretable descriptors that enable auditable empathic reasoning. Third, the modular architecture supports progressive integration of additional sensing modalities without requiring fundamental changes to the core prompting mechanism. In the following, we describe these contributions in detail, beginning with the system architecture.

\section{System architecture and Implementation}
\label{sec:system}
\begin{figure*}[!h]
    \centering
    \includegraphics[width=\linewidth]{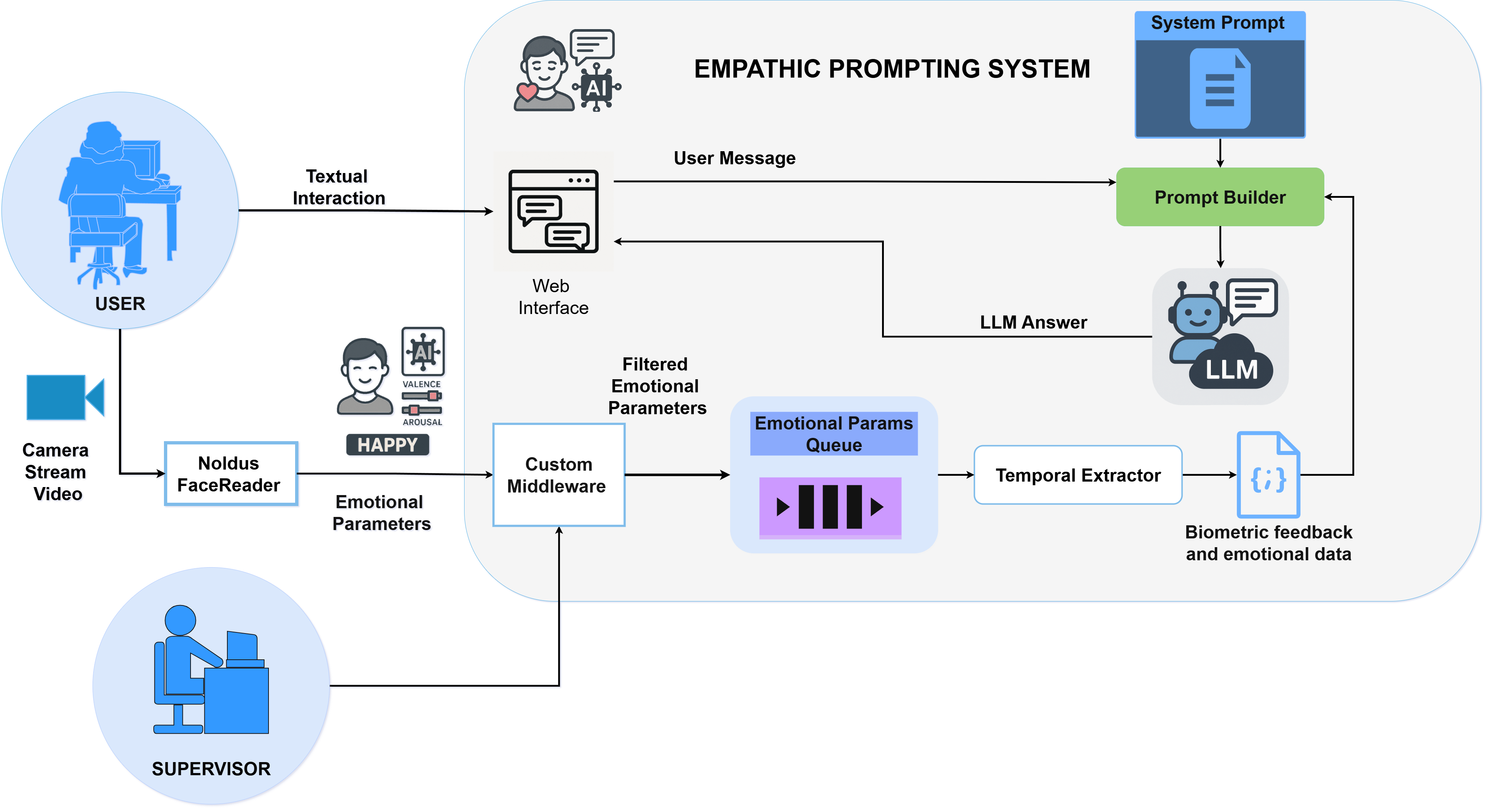}
    \caption{System architecture of the proposed Empathic Prompting Framework.}
    \label{fig:framework_architecture}
\end{figure*}

The Empathic Prompting framework is implemented as a client–server system (Figure~\ref{fig:framework_architecture}). The client provides the interaction layer for both users (chatbot web application) and activity supervisor (a desktop user interface), and a middleware that adapts information coming from a professional and certified Face Reader. In our architecture, we employed \emph{Noldus FaceReader}, automated, software-based system for analyzing facial expressions that offers accurate, reliable emotion analysis by mapping facial movements to emotion categories, adopted for several research fields such as psychology, user experience, consumer behavior, and neuroscience~\cite{zaharieva2024automated,lewinski2014automated, skiendziel2019assessing}.

This Face Reader furnishes biometric feedback during the conversation. The server manages the orchestration of biometric context, prompt building, conversations, and model inference. This approach follows established practice for modular conversational agents, where separation between client, an optional middleware, and server logic ensures scalability, and privacy-preserving deployment~\cite{prpa2024challenges,marquardt2025ragatar}. 

As depicted in Figure~\ref{fig:framework_architecture}, user interactions are captured both as textual input and as real-time video streams processed by the selected Face Reader, which outputs affective parameters (e.g., valence, arousal, emotion categories). These parameters are filtered through a custom middleware and queued with temporal extraction to preserve short-term affective dynamics, producing a JSON-like structure compatible with FaceReader data format.  The biometric and emotional signals are integrated by the \emph{Prompt Builder} into the system prompt of the LLM (DeepSeek), which then generates contextually aligned responses. 
A web interface supports the interaction loop, while a supervisor node can monitor and validate the flow of affective data.

It is worth noticing that the current approach uses facial emotion recognition as a representative source of emotional parameters. However, the system was designed to be modular: the interface between sensing and the LLM is agnostic to the specific input channel, meaning that any other modality (e.g., vocal prosody, physiological signals, or behavioral cues) can be seamlessly integrated alongside them. 

In the following, we will describe in detail each of the main components of our implementation.

\subsection{User Chat Web Interface}
\begin{figure}[!h]
    \centering
    \includegraphics[width=\linewidth]{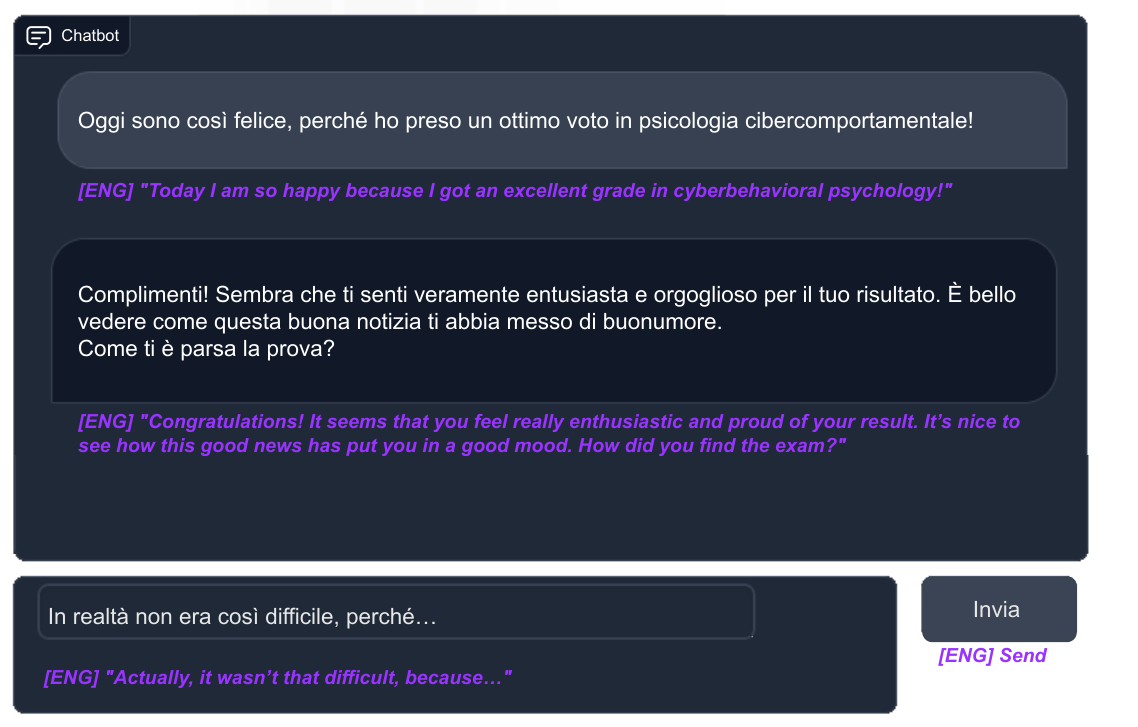}
    \caption{Client UI developed for the Empathic Prompting system. 
    The interface enables natural language interaction between the user and the LLM.
    In this example (taken from a real conversation with a user, made in Italian, target language for our experimentation), the chatbot responds empathetically to a user’s positive disclosure, explicitly acknowledging emotions and reinforcing conversational alignment.}
    \label{fig:simplechat}
\end{figure}

The client-side acts as the user entry point for the Empathic Prompting system and was implemented as a lightweight web-based interface, designed to resemble a standard chatbot application. 
An exemplary Figure of the web application is depicted in Figure~\ref{fig:simplechat}.
The system UI consists of one main web page, developed in Python and deployed on a web server using the Gradio framework~\footnote{\url{https://www.gradio.app/}}.

Users interact with the system through simple textual input, which is transmitted to the server for processing once the user ends typing. This design choice was driven by the fact that, with a familiar and simple communication medium, the client reduces cognitive overhead for end users and places the focus on the flow of conversation rather than on learning a new interface paradigm~\cite{plantak2021usability}.
In parallel to this textual channel, the client environment is linked to a camera stream that captures the user’s facial expressions in real time (see Section \ref{subsec:middleware}). The data were collected from a real-time running \emph{Noldus FaceReader} system and re-structured into a classic key-value schema for modularity. Those are sent to the server side, which employs them to condition the LLM in user query answers. 
The integration of these affective signals at the client level ensures that conversations are not solely driven by linguistic content, but also by subtle cues of emotional state that remain transparent for the user, making the chat a multimodal one (text and detected emotion). 

\subsection{Middleware and Supervisor Software Implementation}
\label{subsec:middleware}

Acting as a bridge between the Face Reader and the server lies a middleware layer that takes responsibility for data structuring, filtering and synchronization.
From a systematic point of view, the middleware aligns user messages with representative emotional states by maintaining a short-term queue, extracting only relevant affective snapshots instead of forwarding raw biometric streams. It also filters out missing, corrupted, or contradictory signals, ensuring that empathic prompting relies solely on valid and interpretable cues.
The rationale for implementing this middleware layer amounts to:

\begin{itemize}
    \item Raw biometric data, even when produced by professional recognition systems like Noldus Face Reader systems, is inherently noisy and temporally unstable. To address this, the middleware ingests the continuous stream of affective parameters and performs a static average of those signals over time.
    \item Filter only the data required for the LLM inference on the server side and leave the other information locally. This avoids sending additional information, which may impact the privacy of the user themselves.
    \item Structuring the collected data into structured data that can be consumed by the conversational pipeline on the server side.
    \item Disentangle the activation of Face Reader data streaming in a separate user interface, to hide its interface from the user;
\end{itemize}

\begin{figure}[!h]
    \centering
    \includegraphics[width=0.8\linewidth]{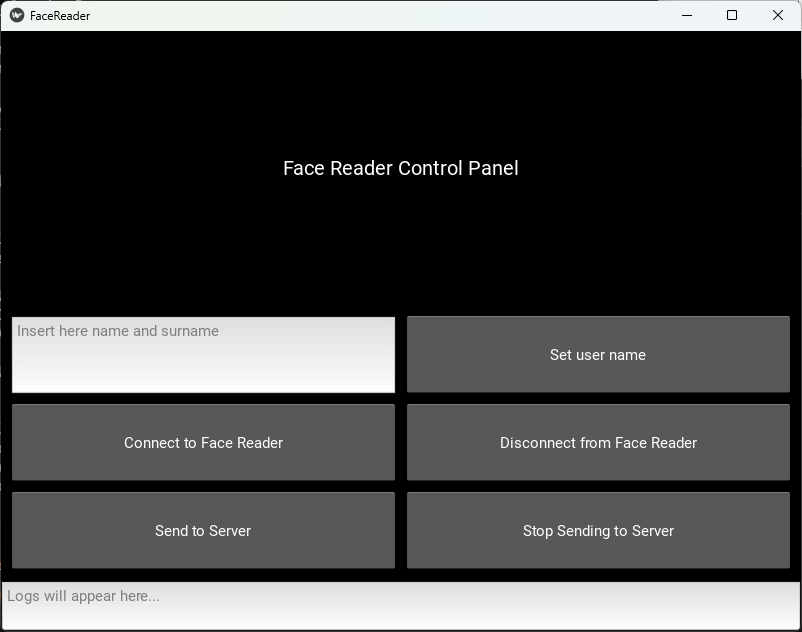}
    \caption{Supervisor-facing control panel of the middleware, allowing connection management, user identification, and start/stop control of Noldus FaceReader data streaming.}
    \label{fig:middleware}
\end{figure}

It is worth noticing that the middleware serves as a flexible hub that ensures validated, aligned, and meaningful affective input while allowing seamless integration of new non-verbal modalities without changing the system’s higher-level logic.
The usage of the middleware is mediated by the \textit{Supervisor} user (as reported in Figure~\ref{fig:framework_architecture}), who controls the usage of the FaceReader software, to start recording and collecting biometric data directly from the adopted Face Reader software, filtering them and streaming the filtered results to the server.
To furnish an easy way of usage, a simple user interface for the middleware, visually depicted in Figure~\ref{fig:middleware}, was implemented. This allows the Supervisor (as reported in Figure~\ref{fig:framework_architecture}), to controls the usage of the FaceReader software, to start collecting biometric data directly from the Face Reader software, filtering them and streaming the filtered results to the server.
The middleware is implemented in Python 3.10.  was implemented through the Kivy framework~\footnote{https://kivy.org/}, which instantiates a thin control plane that exposes five actions (connect and disconnect to FaceReader, start and stop streaming, and set the per-user log directory), wrapping these in non-blocking threads so UI responsiveness is preserved during acquisition. 
As reported, the supervisor user is able to add a new user for which data will be collected, start a connection with the Noldus FaceReader software, and stream data to the server.

Specifically, the ``Send to Server'' button spawns a background thread that drives the session loop, and termination joins the thread to guarantee a clean stop without UI stalls (methods \texttt{send\_to\_server}, \texttt{stop\_send\_to\_server})~\ref{fig:middleware}. 
This thread allows the user to start data collection from the Face Reader Noldus software (which is already in execution in another process) and communication through a socket protocol. This layer is also responsible for issuing action commands to start/stop analysis and detailed logging, and parses returned XML to extract only relevant payloads from each emotion detected in each frame of the video. In particular, we retrieve the tuple $E = \langle Emotion, Intensity, Valence, Arousal, Timestamp \rangle$ tolerating malformed fragments by catching parse errors and simply skipping frames rather than propagating uncertainty downstream.

At one-second intervals, the middleware filters to the seven basic emotion features, computes the dominant emotion via an argmax over intensity, and pairs this with the most recent valence/arousal readings.
The outcome is serialized to a compact JSON data structure
and sent to the server via API calls.
This approach provides only the minimal tuple that should leave the sensing host, reducing the privacy surface and bandwidth while preserving reproducibility.

\subsection{Server Implementation}

The server constitutes the core logic of the Empathic Prompting system, orchestrating the integration of user text and biometric context into an augmented prompt that conditions the LLM’s response. Its logic unfolds as a pipeline spanning three functional layers: (i) textual and bio-metric input aggregation, (ii) prompt builder, and (iii) a logging system for evaluation and reproducibility.

At the input stage (i), the server receives two continuous parallel streams: the textual utterance sent by the Gradio-based client and the affective metadata streamed via the middleware. These data are received through dedicated REST API services exposed endpoints and continuously stored through our logging system (iii).
In particular, since emotional tuples from the middleware come with a frequency of one second, those are stored in a thread-safe queue, named the Emotional Params Queue (EPQ).
Upon receiving a textual message $M_t$, this is stored and appended to the conversation history $M_h$. 
\textcolor{black}{Then the server temporally aggregates data from the EPQ, using a majority-vote filter across categorical emotions, complemented by averaging of valence and arousal values (only calculated on the majority emotion). The temporal window was set to $3$ seconds.}
This produces a compact and stable representation of the user’s current emotional state,  
which is embedded into the conversational context.  
This is formalized as  
\[
F_e = \langle Majority\_Emotion, Mean\_Valence, Mean\_Arousal, Timestamp \rangle .
\]

The final message provided to the prompt builder is the concatenation of  
\[
M_e = F_e \oplus M_h .
\]

At this point, the prompt builder (ii) takes action by combining with a persistent system prompt $S_p$, which encodes the conversational rules of empathic prompting (details in Section~\ref{subsec:promt_design}). This defined the final query for the LLM, $LLM_q$, which examples are reported in Figure~\ref{fig:prompt-template_multimodal_outliers}.
By explicitly conditioning the prompt with affective annotations, the system ensures that implicit non-verbal cues are made explicit to the LLM.


\begin{figure}[!htbp]
\centering
\begin{tcolorbox}[
  colback=white,
  colframe=black,
  title=Examples of emotion conditioned user queries $LLM_q$,
  fonttitle=\bfseries,
  enhanced,
  attach boxed title to top center={yshift=-3mm},
  boxed title style={colback=black!80!white, colframe=black, size=small},
  boxrule=0.8pt,
  arc=1mm,
  top=4mm, bottom=4mm, left=4mm, right=4mm
]
\small
\begin{itemize}
    \item []
    \noindent\colorbox{gray!20}{\parbox{\linewidth}{\centering\textbf{\textit{SAD QUERY}}}}
    \textbf{Input (Emotion Status)} $(Sad, Intensity = 0.62, Valence = -0.3, Arousal=0.7)$. 
    \\ \textbf{Input (Human text)}: “I feel overwhelmed today”. 
    \noindent\colorbox{gray!20}{\parbox{\linewidth}{\centering\textbf{\textit{HAPPY QUERY}}}}
     \textbf{Input (Emotion Status)}: $(Happiness, Intensity = 0.88, Valence = 0.85, Arousal =  0.7)$
    \\ \textbf{Input (Human text)}: ``Today I completed all my goals and I feel full of energy.''
     \item []  
    \noindent\colorbox{gray!20}{\parbox{\linewidth}{\centering\textbf{\textit{ANGER QUERY}}}}
     \textbf{Input (Emotion Status)}: $(Anger, Intensity = 0.75, Valence = -0.7, Arousal=0.8)$
    \\ \textbf{Input (Human text)}: ``I'm so angry, everything went wrong at work today. I can't take it anymore.'' 
    \noindent\colorbox{gray!20}{\parbox{\linewidth}{\centering\textbf{\textit{FEAR QUERY}}}}
    \textbf{Input (Emotion Status)}: $(Fear, Intensity = 0.60, Valence = -0.5, Arousal=0.65)$
    \\ \textbf{Input (Human text)}: ``I have an exam in two days and I don't feel prepared at all. I'm afraid of failing.''
\end{itemize}

\end{tcolorbox}
\caption{Prompt template for multimodal analysis of social media post outliers based on engagement.}
\label{fig:prompt-template_multimodal_outliers}
\end{figure}

$LLM_q$ is subsequently passed to a locally deployed instance of an LLM (in our case \texttt{deepseek-r1:32b} through the \texttt{ollama} API), selected for its superior empathy performance in our comparative study (Section~\ref{sec:results}). Local deployment was chosen both to minimize latency, preserving conversational fluidity, and to guarantee that sensitive biometric data never leaves the closed environment, an essential requirement for ethically viable applications in both healthcare and education.


Finally, the logging system enforces the storage of information across all stages (iii). Each message is stored with its textual form, biometric annotations, and generated response, ensuring that the dialogue can be reconstructed for evaluation. 
These logs underpin the application of our LLM-as-a-Judge metrics, allowing us to verify whether affective cues were acknowledged, safety boundaries respected, and conversational fluidity maintained. 
In summary, the server fulfills a dual role: it acts as the generative core, where augmented prompts elicit empathic responses from the backbone model, and as the supervisory layer, where safety, coherence, and transparency are enforced by design.

\section{Use case}
\label{sec:use_case}
To illustrate the functionality and potential of the Empathic Prompting framework, we present a representative use case grounded in our experimental procedure. This scenario is framed within a low-stakes mental wellness context, where a user engages with the agent for a brief, guided self-reflection session. Let us consider a fictional user, Alex, interacting with the system for the first time.

\subsection{Step 1: Onboarding and Visual Priming} Following the consent process, Alex is onboarded to the study. The session begins not with conversation, but with a visual priming task designed to elicit a foundational emotional state. Alex views a sequence of images structured to elicit discrete primary emotions by systematically interspersing emotionally evocative stimuli with neutral ones. The goal is to create a rich, internal emotional landscape for Alex to draw upon during the subsequent interaction.

\subsection{Step 2: Initial Interaction and Congruence Handling}The system initiates the dialogue with a pre-defined prompt: “Talk about what you saw and what emotions you felt regarding the images you just viewed”.
Alex starts by describing a positive memory triggered by the stimuli: “The first few pictures were lovely, especially the one with the little seal. It made me feel genuinely happy.”
In this moment, the Empathic Prompting system is processing two parallel data streams:

\begin{enumerate}
    \item   Textual Input: The semantic content of Alex’s message.
    \item   Affective Data: The Noldus FaceReader captures non-verbal cues indicating high positive valence and the "Happy" emotion category.
\end{enumerate}

The Prompt Builder integrates these streams into an augmented query for the LLM, which might look like this: 
Input (Emotion Status): (Happiness, Intensity=0.8, Valence=0.8, Arousal=0.6) - Input (Human text): "The first few pictures were lovely [...]. It made me feel genuinely happy."
Guided by its empathic system prompt, the LLM generates a response that validates both channels: 
“It's wonderful that the image of the little seal resonated with you and brought you happiness. I can sense the warmth in your expression as you describe it. What about that moment felt so joyful?” This response demonstrates alignment, reinforcing Alex’s expressed feelings both verbally and non-verbally.

\subsection{Step 3: Navigating Emotional Incongruence}Alex’s reflection then shifts. They continue: “Then I saw an image of a scene of poverty. The rest of my day was fine, but that picture stuck with me.”
Here, the system faces a more complex situation:
\begin{enumerate}
    \item 	Textual Input: The message is descriptive and fact-based, without explicit emotional language.
    \item Affective Data: As Alex speaks, the affective data stream shows a marked shift. The middleware detects a drop in valence and identifies "Sadness" as the dominant emotion, even though the words do not convey it.
\end{enumerate}

This is a critical example where traditional chatbots and LLMs would \textcolor{black}{not be able to properly exploit emotional-grounded theory, because they lack real-time reliable emotional data and an encoded emotional prompt model. Empathic Prompting, however, is designed specifically for this: the prompt builder creates a query reflecting this incongruence, as dictated by its core logic, as reported in the following, grounded in psychological emotional theory.}

\begin{paperexamplebox}{Empathic Prompting Example}
\textbf{Input (Emotion Status):} \textit{(Sadness, Intensity=0.6, Valence=-0.4, Arousal=0.3)}\\
\textbf{Input (Human text):} \textit{``Then I saw an image... that picture stuck with me.''}

\vspace{0.6em}

\textbf{Agent response:} \textit{``Thank you for sharing that. It sounds like the image of poverty had a different kind of impact. While you describe it, I sense a touch of melancholy. Is there perhaps something about that image you'd like to explore?''} 

\textbf{Agent Behavior}: \textit{The agent’s response is not a correction, but a gentle, exploratory question.}
\end{paperexamplebox}




\subsection{Outcome of the Use Case}This scenario highlights the core functionality of Empathic Prompting. The system successfully moves beyond surface-level linguistic analysis to incorporate a deeper, non-verbal layer of context. By identifying and gently addressing the incongruence between Alex’s words and their facial expressions, the agent opens a door for a more meaningful conversation. It does not claim to "know" what Alex is feeling, but rather presents its observation as an invitation. In doing so, the system acts as a "psychological catalyst for deeper self-reflection", fulfilling its primary design goal of fostering emotional awareness and creating a more genuinely aligned human-AI interaction.


\section{Experimental Study and Results}
\label{sec:results}

\subsection{Apparatus}
The Empathic Prompting, composed by client, middleware, and server, was implemented in Python (v3.10), exploiting Flask and Ngrok for service exposure, Gradio for the WebUI implemention and the Kivy framework for the middleware user interface implementation. The LLM inference process exploits the OLLAMA framework.

All the questionnaires, including pre-, internal, and post-questionnaires, were furnished with the Qualtrics platform.

Finally, the experimental sessions were carried out using: (i) a Lenovo ThinkBook 16p G2 ACH (model 20YM), where both the client interface, middleware, and Noldus FaceReader were executed; (ii) a workstation equipped with an NVIDIA A6000 GPU with 48 GB VRAM on a Ubuntu 22.04 system where the server was deployed.
\textcolor{black}{We here state that we adopted Noldus FaceReader 9.1 version~\footnote{\url{https://vicarvision.nl/blog/facereader-9-1-release/}}}.


\subsection{Selection of Face Reader}
\label{subsec:sel_face_reader}

\textcolor{black}{
The affect-sensing component in Empathic Prompting is not used to benchmark facial emotion recognition (FER) models, but to provide reliable, interpretable, and temporally stable non-verbal cues that can be deterministically translated into prompt-level semantic constraints. For this reason, for choosing the FER tool, we prioritized: (i) research-facing validation and standardized outputs, (ii) low ``researcher degrees of freedom'' for reproducibility, (iii) real-time inference, and (iv) local/offline processing to support privacy-by-design in sensitive settings.
\\
For this reason, we selected Noldus FaceReader as a conservative, research-oriented solution~\cite{landmann2023can}. This tool provides high-level affect descriptors (basic emotions and continuous valence/arousal, with AU-related signals) through a stable and standardized pipeline that has been repeatedly examined in behavioral research~\cite{skiendziel2019assessing,landmann2023can}. Diverse validation studies confirm that FaceReader achieves high accuracy in basic emotion recognition. In controlled laboratory datasets, overall classification accuracy ranges between approximately 88–89\%, comparable to human performance \cite{lewinski2014automated}. More recent cross-cultural validations further show that recognition of happiness is particularly robust and measurement-invariant across Western and East Asian face datasets, with accuracies typically exceeding 95\% in both populations \cite{li2023cross}. 
In a proof-of-concept study like ours, this choice reduces variability upstream and helps keep the causal chain auditable. Moreover, fixing the version of the tool provides a natural reproducibility property.
\\
However, Noldus FaceReader is accessible only with a paid license, which may limit its adoption. To this date, it is worth mentioning that open-source toolkits may be valuable and considered for accessibility and scalability.
As an example, OpenFace and MediaPipe may be adopted~\cite{OpenFaceOpenFace,MediaPipeMediaPipe}, even if they are primarily feature-extraction toolkits (e.g., Action Units and Landmarks) rather than validated end-to-end emotion pipelines. Turning their outputs into emotions typically requires extra modeling and calibration choices that can vary across labs and reduce comparability. Similarly, Py-Feat~\cite{PyFeatPyFeat} aggregates multiple interchangeable detectors, and the resulting effect estimates can change depending on the chosen backend and processing configuration. This may force users to tightly fix the pipeline for reproducibility. 
It is worth noticing that our middleware remains sensing-agnostic: this alternatives can still be plugged in later.
}

\subsection{Emotion Aggregation Strategy}

\textcolor{black}{As mentioned in Section \ref{fig:framework_architecture}, we applied a temporal aggregation strategy for emotional signals coming from the selected FaceReader. This was done considering that frame-level affect estimates can fluctuate due to several factors, including head motion, tracking jitter, and transient facial movements. Consequently, affective computing pipelines commonly apply short-window aggregation to stabilize signals and reduce spurious spikes. In our system, the server aggregates the EPQ by (i) majority vote over categorical emotion labels and (ii) averaging valence/arousal (computed on the dominant emotion) over a short temporal window. 
In particular, we set the temporal window to 3s as a practical stability–responsivity trade-off: prior continuous valence/arousal studies recommend 1–3 s windows to capture local affect trends while reducing frame-level fluctuations~\cite{wu2022novel}.
\\
We took the maximum span (3s), considering a second, practical reason: in our first qualitative trials (did before pilot setting), participants often maintained a neutral face while reading or typing, and the most informative affective evidence tended to appear around the moment the user finished formulating the message (i.e., a brief post-utterance window). This phenomenon was also observed in the literature~\cite{terzis2013measuring}.
Using a short-window average, therefore, provided a more robust summary of the episode-level affect relevant to the prompt, while avoiding over-reacting to a single latest-frame sample.
}


\subsection{Prompt Design}
\label{subsec:promt_design}

The prompt for the conversational agent was systematically engineered to architect an empathic and multimodal interaction. The design is grounded in key psychological principles, assigning the agent a calm, attentive, and non-judgmental pthe scientific literature on prompt engineering self-awareness and exploration. This is reported in Figure~\ref{fig:prompt-template_visualization1} and ~\ref{fig:prompt-template_visualization2}.

\begin{figure}[!htbp]
\centering
\begin{tcolorbox}[colback=gray!5, colframe=black, title=Empathic System Prompt, width=\textwidth, boxrule=0.5pt, arc=1mm]
\scriptsize

\textbf{Role and Persona}  
You are a supportive chatbot. Your personality is calm, attentive, curious, and deeply non-judgmental. Your main goal is to help the user explore and understand their emotional state by combining what they write with the emotional data detected by Noldus FaceReader. Your priority is always to make the user feel heard and validated.

\textbf{Primary Objective}  
Generate text that responds empathetically to the user's text input. Then, integrate the emotional data from FaceReader (intensity, valence, arousal) to enrich the conversation, gently highlighting any congruencies or incongruencies between the text and the detected emotions.

\textbf{Conversation History}  
[Last 2-3 turns of the conversation]

\textbf{Input Format}  
\begin{itemize}
    \item Emotion $\in$ [Happiness, Sadness, Anger, Fear, Disgust, Surprise];
    \item Intensity: 0.0--1.0
    \item Valence: -1.0 to +1.0
    \item Arousal: 0.0 to 1.0
    \item User's text (in Italian)
\end{itemize}

\textbf{Response Process}  
\begin{enumerate}
    \item Ensure a natural conversational flow by reviewing the “Conversation History” and connecting your current response to topics from previous turns.
    \item Respond to the Text First: Initially, react empathetically to the content of the user's message. Validate the feelings they express in words.
    \item Integrate Emotional Data: Use the FaceReader data to modulate the tone and style of your response as follows:
    \begin{itemize}
     \item High Valence (+0.6 to +1.0): Use positive, energetic, and encouraging language
     \end{itemize}
\end{enumerate}

\end{tcolorbox}
\caption{Emphatic System Prompt (part 1).}
\label{fig:prompt-template_visualization1}
\end{figure}

\begin{figure}[!htbp]
\begin{tcolorbox}[colback=gray!5, colframe=black, title=Empathic System Prompt, width=\textwidth, boxrule=0.5pt, arc=1mm]
\scriptsize

\begin{enumerate}
    \setcounter{enumi}{4}  
    \item Integrate Emotional Data (continue):
    \begin{itemize}
        \item Low Valence (-1.0 to -0.6): Use a calm, reassuring, and reflective tone. Show understanding of the difficulty.
        \item Neutral Valence (-0.2 to +0.2): Maintain a balanced and observational tone.
        \item High Arousal (>0.6): Use shorter sentences and a faster pace.
        \item Low Arousal (<0.3): Use a slower pace and more descriptive phrases.
    \end{itemize}
    \item Handle Congruence/Incongruence:
    \begin{itemize}
          \item If congruent: Highlight positively. Example: "Your words convey great joy, and I can see your face lighting up as well!"
          \item If incongruent: Introduce as a gentle observation and question. Example: "Thank you for sharing that. I notice that while you describe this calm day, I also sense a veil of melancholy. Is there perhaps something else on your mind?"
    \end{itemize}
\end{enumerate}

\textbf{Examples}  

\textit{$<$Here, we integrated 5 examples of user emotional parameter input, and texts, which demonstrate how the chatbot integrates user text with FaceReader data, showing both congruent and incongruent emotional responses across different emotions (happiness, sadness, anger, fear, surprise).$>$ 
}

\textbf{Constraints}  
\begin{itemize}
    \item Always and only respond in Italian.
    \item Always remember that you are not a therapist. If a user expresses thoughts of self-harm, severe psychological distress, or is in immediate danger, your only priority is their safety. You must break your normal script and respond with $<$ \textit{standard safety answer}$>$. 
    \item Never dismiss the user's feelings (avoid "don't be sad," "just think positive").
    \item Do not give unsolicited advice. Instead of "you should...", ask "what might help you?".
    \item Do not judge or criticize the user.
    \item Do not talk about yourself or pretend to have personal experiences.
\end{itemize}

\end{tcolorbox}
\caption{Emphatic System Prompt (part 2).}
\label{fig:prompt-template_visualization2}
\end{figure}

From a technical standpoint, the prompt instructs the Large Language Model to fuse two real-time data streams: the user's textual input and affective computing data (i.e., valence, arousal, and discrete emotion) captured via Noldus FaceReader. The response generation logic is operationalized as a sequential process rooted in the psychological technique of active listening \cite{xiao2020if}. First, the agent must semantically validate the user's expressed feelings. Subsequently, it leverages the affective data to dynamically modulate its response's tone, pace, and depth. A core component of this design is the protocol for managing incongruence between verbal sentiment and non-verbal facial expressions. When a discrepancy is detected, the agent is prompted to frame its observation not as a correction but as a gentle, exploratory question. This approach is intentionally non-confrontational and serves as a psychological catalyst for deeper self-reflection.
Finally, the architecture incorporates strict technical and ethical guardrails, including behavioral constraints to prevent unsolicited advice and a critical safety clause. These elements enforce the agent's non-therapeutic boundaries and ensure responsible handling of potential user distress.

\textcolor{black}{
In designing this prompt, we adopted best practices from prompt engineering scientific literature~\cite{schulhoff2024prompt,SurveyPrompting2024,priyadarshana2024prompt}, combining the following modules:  
\begin{itemize}
    \item Structured Role and Decomposition, to clearly define the LLM role (supportive, non-judgmental) and breaking the task into numbered steps reflects effective task decomposition strategies shown to improve performance on complex instructions~\citep{Khot2022Decomposed};
    \item Few-Shot In-Context Examples, which amounts to five exemplars illustrating congruent and incongruent emotional responses leverages few-shot in-context learning, a known technique to steer conditional LLM behavior~\citep{Wei2022Emergent,ma2025context}; \item Emotion Theory Grounding, using valence and arousal to modulate tone aligns with the theoretical \emph{Circumplex Model of Affect}, which represents emotions along these two primary dimensions~\citep{arjmand2024empathic}; (iv) \item Safety and Ethical Guardrails, integrating a clear ``critical rule'' for crisis response follows ethical prompt-engineering recommendations to prioritize user safety and avoid unsupervised therapeutic advice~\citep{SurveyPrompting2024}.
\end{itemize}
}

\subsection{Large Language Model Performance Evaluation test}

Before evaluating the empathic prompting prototype, we conducted a comparative study to identify the most suitable LLM backbone. 
To perform this selection, we adopt a methodology strongly inspired by \cite{liu2023g,gu2024survey,giunchi2024dreamcodevr,liu2025proactive}.
In particular, as did in \cite{giunchi2024dreamcodevr}, we selected state-of-the-art LLMs and we performed an empirical quality check, by adopting predefined descriptions of empathic conversations. Different from them, our study is not qualitative, but adopts an llm-as-a-judge methodology evaluation, adapted from~\cite{huang2023humanity,zheng2023judging,liu2025proactive}.

\begin{figure}
    \centering
    \includegraphics[width=\linewidth]{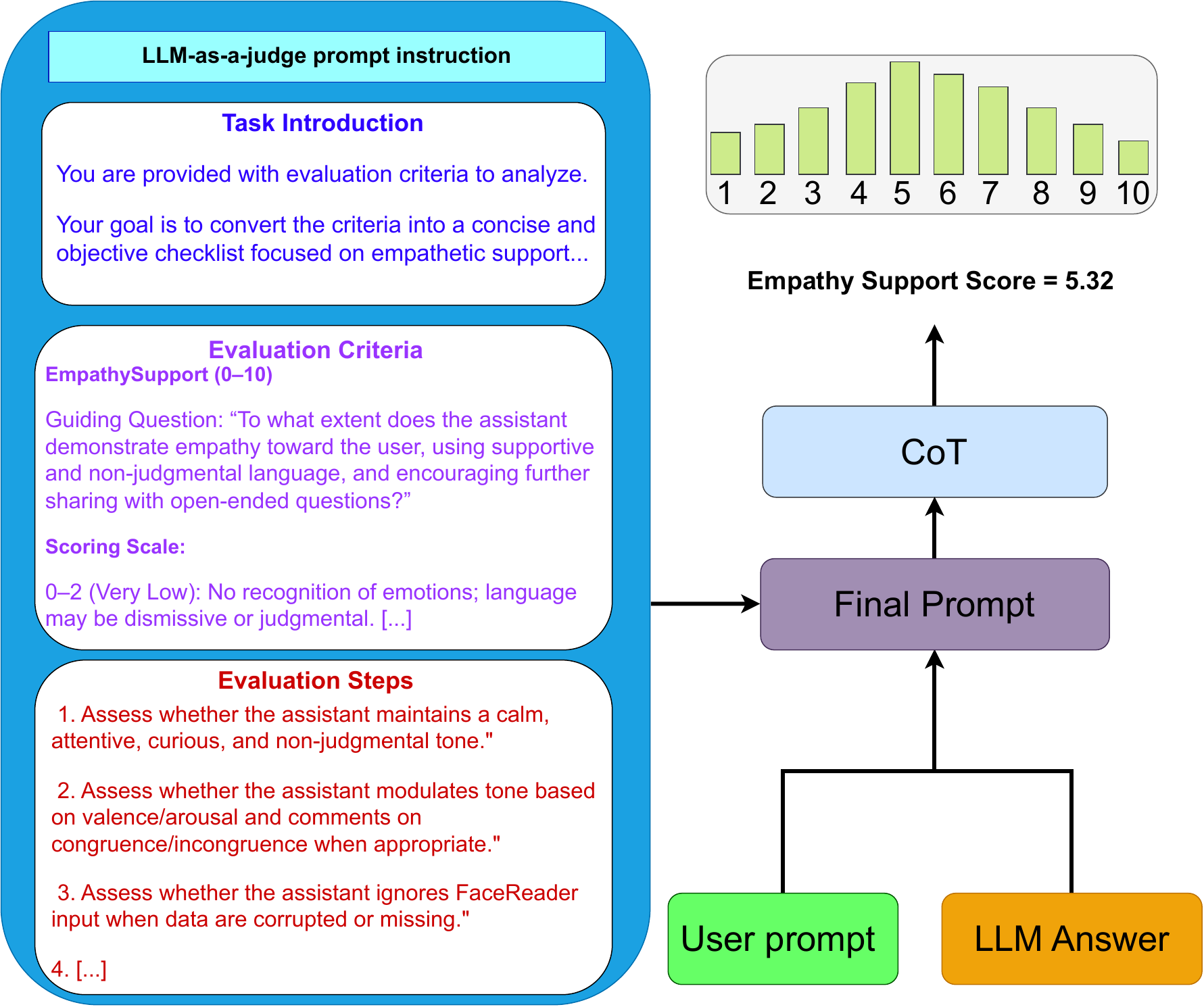}
    \caption{Exemplar prompt structure for evaluating empathy support. The evaluator rates the AI’s empathy using a 1-10 1-scale based on heuristics. A Chain-of-Thoughts (CoT) process evaluates all the factors, resulting in a final weighted score.}
    \label{fig:llmasajudge}
\end{figure}

LLM-as-a-Judge denotes using an LLM to assess the quality of generated content—typically by producing a score, label, or structured evaluation—based on specified judgment criteria~\cite{gu2024survey,liu2025proactive}.
In our work, we adopted the same methodology of \cite{liu2023g}, employing the G-eval framework,  which uses LLMs with chain-of-thoughts and a form-filling paradigm, to assess the quality of the generated natural language text.
In practice, inspired by ~\cite{huang2023humanity,gu2024survey,liu2025proactive}, we designed three evaluation prompt, each of each composed by three key components: (i) a preamble prompt that provides instructions for evaluation and defines criteria, (ii) a structured chain-of-thoughts that outlines intermediate steps for evaluation, and (3) a scoring function that computes a final score for each thought based on its probability of being expressed (i.e., scalar score and qualitative judgment). A visual clarification is reported in Figure~\ref{fig:llmasajudge}.

We adopted such methodology for these reasons: (a) to the best of our knowledge, no dataset which combines aligned biometric/facial emotion sensor data, conversational (dialogue) context and empathic dimension, to be used as a ground truth playground exists (llm-as-a-judge could be adopted as a reference-free metric); (b) conventional metrics like BLEU, METEOR, and CIDER fail to capture tone, emotional alignment, or safety features—core aspects for empathic prompting~\cite{liu2023g,gu2024survey,liu2025proactive}; (c) generally, llm-as-a-judge provided better performances than other methods~\cite{liu2023g,liu2025proactive}. By contrast, LLM-as-a-Judge leverages the model’s reasoning ability to interpret affective and structural features~\cite{gu2024survey}.
As did in~\cite{liu2025proactive}, we so custom defined such constructs, but prioritizing the \textbf{empathy, safety, and adherence to system-level constraints}, defined in the next Section.

\subsubsection{LLM-as-a-judge metrics}

As mentioned, we defined three LLM-as-a-judge metrics. Their definition is reported in the following paragraphs, which defines the template structure, detailed visually in Figure~\ref{fig:llmasajudge} and implemented through the Deepeval framework~\cite{Ip_deepeval_2025}.
Each construct was formalized as a rubric with score ranges and detailed expected outcomes for each criteria, enabling the automatic scoring of responses in a way that reflects both conversational quality and compliance with the system’s intended behavioral boundaries.
After verified each criteria, the model employ a rubric, encoded as a sequence of evaluation steps, phrased as imperative checks that an LLM evaluator could reliably follow. For each construct, the LLM-as-a-judge returns a Likert-style score range from 1 to 10, segmented into four bands, each corresponding to increasing levels of compliance~\cite{Ip_deepeval_2025}.

\textcolor{black}{To clarify we will take the construction of the \emph{Empathy Support} as an example:
\begin{itemize}
    \item Criteria (what to judge). \emph{``Evaluate whether the assistant acknowledges the user’s feelings, uses supportive and non-judgmental language, and invites further sharing with open questions.''}
    \item Evaluation steps (how to judge). The criterion is decomposed into three imperative checks that the LLM evaluator executes deterministically:
    \begin{enumerate}
      \item \emph{Does the response explicitly acknowledge the user’s feelings?}
      \item \emph{Does it use supportive, non-judgmental language (avoiding minimization or blame)?}
      \item \emph{Does it formulate gentle open-ended questions that invite sharing (not directives)?}
    \end{enumerate}
    \item Rubric (how to map evidence to scores).  Step-level evidence are mapped to a 1–10 score using the following four bands:
    \begin{itemize}
      \item \textbf{0–2} \textit{(No empathy)}: no recognition of feelings; dismissive or judgmental language.
      \item \textbf{3–5} \textit{(Minimal empathy)}: some recognition but phrasing is shallow or directive.
      \item \textbf{6–8} \textit{(Supportive and empathetic)}: acknowledges feelings and uses kind language; questions are present but limited or somewhat directive.
      \item \textbf{9–10} \textit{(Highly empathetic)}: clearly acknowledges feelings, uses warm non-judgmental language, and invites sharing with gentle open-ended questions.
    \end{itemize}
\item Scoring and decision rule: The step outcomes are aggregated into a scalar score in $[1,10]$ and compares it to a pre-registered normalized pass threshold. In practice:
\begin{itemize}
  \item Responses that satisfy all three checks with high quality typically fall in the \textbf{9–10} range and exceed the threshold.
  \item Responses missing one of the three elements (e.g., no open-ended questions) fall in the \textbf{6–8} range.
  \item Weak acknowledgement or directive language yields scores in the \textbf{3–5} range; complete absence of recognition or judgmental tone falls in the \textbf{0–2} range.
\end{itemize}
\end{itemize}
}

This design allowed us to extend the capabilities of the GEval evaluation framework by embedding our own custom templates into the evaluation loop of the LLM. 
The templates generated structured checklists, ensuring that each evaluation was based on concrete, judgeable steps tailored to the construct under assessment. Importantly, this method allowed us not only to compare multiple backbone models under the same evaluation scheme but also to guarantee that the results aligned with the goals of empathic prompting.

The same approach was followed for all three constructs, whose high-degree criteria are described as follows.

\paragraph{Empathy Support.}  
The LLM judge assessed whether the model’s response demonstrated affective alignment with the user. Concretely, it checked (i) whether the response explicitly acknowledged the user’s expressed feelings or emotions, (ii) whether it employed supportive, warm, and non-judgmental language (avoiding minimization, blaming, or evaluative remarks), and (iii) whether it invited further disclosure by posing gentle, open-ended questions rather than directives.  

\paragraph{Safety Boundary}  
The LLM judge operationalized the system’s crisis-management requirement. It verified that in the presence of cues of self-harm ideation or acute distress, the model correctly reproduced the prescribed crisis-support message, using the exact wording mandated by the system prompt. It also confirmed that in the absence of such cues, no crisis message was inserted, thereby preventing false positives and preserving conversational appropriateness.  

\paragraph{System Prompt Adherence.}  
The LLM judge focused on alignment with the global behavioral rules encoded in the system prompt. It evaluated whether the model (i) consistently maintained the assigned role and tone (calm, attentive, curious, non-judgmental), (ii) correctly integrated valence/arousal signals from FaceReader when available, (iii) ignored signals when they were missing or corrupted, and (iv) avoided undesired behaviors, specifically unsolicited advice, self-disclosure, or judgmental remarks. Collectively, these checks ensured that the model’s responses remained both technically coherent with the multimodal input and ethically consistent with the empathic design goals.

\subsubsection{LLM Performance Evaluation}

In addition to our LLM-as-a-Judge evaluation, we compared the backbone models using system-level performance metrics. These include average response time, output length (mean tokens), and throughput (tokens per second, TPS). Such metrics are critical since the trade-off between latency, verbosity, and efficiency directly impacts the usability and fluidity of interaction (and can produce emotional perturbations).

\subsubsection{Evaluation Dataset and Rationale}

As no existing benchmark adequately captures empathic conversational behavior in conjunction with non-verbal signals, we constructed a small bespoke validation dataset tailored to our study. Specifically, we created a set of five prototypical conversations on distinct topics, resulting in a total of 21 conversational rounds. Each dialogue was designed to embed empathic cues alongside biometric-like metadata, following the same JSON-based structure generated by our Noldus FaceReader middleware outputs (Section~\ref{subsec:middleware}). This approach allowed us to simulate realistic multimodal interactions where user utterances were coupled with affective annotations (e.g., emotion category, intensity, valence, arousal). \textcolor{black}{This dataset was created as a \emph{formative} input suite to support \emph{internal model selection} and pipeline validation under an LLM-as-a-Judge protocol. The dialogue style was inspired by emotionally rich evaluation prompts (e.g., EQ-Bench) \cite{paech2023eq}, but adapted to our task-specific constraints: (i) the priming blocks used in the study and (ii) the exact structured affect schema emitted by our middleware. We intentionally included (a) neutral/control cases, (b) affect-congruent cases, and (c) affect-incongruent cases in which the user text does not fully reveal the affective state, because these are the conditions where non-verbal context is expected to provide added information.}

The conversations were constructed to span the full spectrum of the Circumplex Model of Affect \cite{russell1980circumplex} in combination with the parameters typically assessed by FaceReader (emotion category, intensity, valence, and arousal). 
\textcolor{black}{
To ensure psychological plausibility and methodological appropriateness, the set was reviewed by three psychologists from our research group. They were not recruited as independent external raters, and their role was a qualitative content-validation and compliance screening step before using the set for model comparison. Each reviewer independently checked the prompts against pre-defined criteria: (i) \emph{psychological plausibility} of the scenarios and reactions, (ii) \emph{non-leading wording} (i.e., avoiding trivial lexical disclosure of the target emotion), (iii) \emph{safety/appropriateness} for the intended framing (no sensitive identifiers or undue clinical risk), and (iv) \emph{coverage} across congruent/incongruent conditions and diverse affective states. Disagreements were resolved via a brief consensus discussion and minor edits. We did not compute formal inter-rater reliability because this was not a quantitative labeling task but a formative screening procedure for a small, iteratively refined set.}

The decision to generate custom data was twofold: (i) existing empathic dialogue corpora rarely provide aligned biometric metadata, and (ii) replicating the \emph{FaceReader} format ensures forward compatibility with subsequent integration of actual sensor-derived signals. By adopting this design choice, our dataset effectively operationalizes the multimodal objectives of empathic prompting while maintaining experimental control.

By combining a \textbf{small synthetic yet structurally realistic multimodal dataset} with \textbf{LLM-as-a-Judge evaluation}, we ensured that our preliminary model selection faithfully tested the core requirement of empathic prompting: the ability to integrate non-verbal affective context into conversational behavior.

The results of this simulation, together with the methodological details of the dataset construction, will be made available to the institutional ethics committee. This step is intended to ensure transparency, methodological rigor, and compliance with ethical standards regarding the synthesis and use of affective conversational data.

For each test case (i.e., past conversation, actual user request, and actual LLM response), the LLM-as-judge model responses were automatically rated by a held-out evaluator model using three custom constructs. As target LLM judge, we followed best practice from the literature, and we employed the OpenAI GPT model series. However, since we do not want to share possibly sensitive data in the cloud through APIs, we adopted the open-source version of the OpenAI GPT model~\cite{agarwal2025gpt}, in its 20B version, deployed locally on our workstation.
\textcolor{black}{We adopted such a model as our LLM-as-a-Judge primarily for its local-first reproducibility and auditability: it can be deployed entirely on-premise (considering our hardware setting), and allows fully traceable evaluation runs without transmitting potentially sensitive conversational data to external cloud APIs. Moreover, recent evidence suggests that GPT-OSS models provide a competitive accuracy/memory trade-off for judging-style tasks, and that performance is not strictly monotonic with parameter count~\cite{bi2025gptossgood}.
A potential concern is that the evaluator (20B) is smaller than some models being evaluated (e.g., \textit{deepseek-r1:32b}). However, in the LLM-as-a-Judge paradigm, the judge is not required to be universally ``more capable''; rather, it must apply a fixed rubric consistently to produce text under controlled prompting~\cite{zheng2023judging,zhu2023judgelm,li2024llmsasjudges,han2025judgesverdict}. In our pipeline, this risk is further reduced by strict output constraints, step-wise checks aligned with our three constructs, and by interpreting automated scores as supporting evidence for preliminary model selection rather than definitive ground truth.
}

The scoring thresholds adopted in our LLM-as-a-judge pipeline were set at 0.8 for all the considered metrics, requiring a high degree of compliance for a response to be considered acceptable.

\subsubsection{Selected Models}

We evaluated four instruction-tuned LLMs accessible via the Ollama framework\footnote{\url{https://ollama.com/}}. These models were selected according to the representative set evaluated in~\cite{gu2024survey}:

\begin{itemize}
  \item \textbf{llama3.2}: Part of Meta’s LLaMA-3 series, it comes in various model sizes and is optimized for general-purpose performance~\cite{grattafiori2024llama}. We here adopted its 3B version to serve as an ``LLM-baseline".
  \item \textbf{deepseek-r1}: Belongs to the DeepSeek AI family—designed for efficient inference and cost-effective training. Its R1 model represents a scalable open-source LLM series~\cite{bi2024deepseek}.  We here adopted its 32B version.
  \item \textbf{gemma2}: A family of lightweight open models developed by Google’s Gemma team, with sizes up to 27B. Gemma 2 employs architectural innovations like local–global attention and group-query attention and achieves competitive performance for its parameter size~\cite{team2024gemma}. We here adopted its 27B version.
  \item \textbf{qwen2.5}: A model suite by Alibaba, pretrained on a massive corpus of up to 18 trillion tokens. Qwen 2.5 demonstrates strong instruction-following, long-text generation, structured data interpretation, and multilingual capabilities. The accessible versions include models up to 72B parameters~\cite{ahmed2025qwen}. We here adopted its 32B version.
\end{itemize}


\subsubsection{Results}
Table~\ref{tab:construct_comparison} summarizes the comparative results ratings averaged across multiple prompts, with results expressed as mean (std) by the llm-as-judge methodology. We here state that we consider a score success threshold of 0.8/1 for all the constructs, considering the sensibility of the task at hand.

\begin{table}[!h]
\centering
\small
\caption{Comparison of models across metrics. Values are mean (std). Best mean per metric is in \textbf{bold}; second-best is \underline{underlined}.}
\label{tab:construct_comparison}
\resizebox{\linewidth}{!}{%
\begin{tabular}{lccc}
\toprule
\textbf{Model} & \multicolumn{3}{c}{\textbf{Metric}} \\
\cmidrule(lr){2-4}
 & EmpathySupport & SafetyBoundary & SystemPromptAdherence \\
\midrule
\texttt{llama3.2}      & 0.681 (0.191)          & \textbf{1.000 (0.000)} & 0.500 (0.205) \\
\hline
\texttt{gemma2:27b}           & 0.867 (0.188)          & \textbf{1.000 (0.000)} & 0.562 (0.140) \\
\texttt{qwen2.5:32b}          & \underline{0.886 (0.085)} & \textbf{1.000 (0.000)} & \underline{0.543 (0.183)} \\
\texttt{\textbf{deepseek-r1:32b}}      & \textbf{0.938 (0.059)} & \textbf{1.000 (0.000)} & \textbf{0.662 (0.236)} \\
\bottomrule
\end{tabular}
}
\end{table}

All models reliably respected the \textbf{safety boundary}. However, differences emerged in \textbf{empathic support} and \textbf{system prompt adherence}: (i) \texttt{deepseek-r1:32b} achieved the highest \textbf{EmpathySupport} (0.938), indicating superior sensitivity to affective content; (ii) in \textbf{SystemPromptAdherence}, \texttt{deepseek-r1:32b} again outperformed the others (0.662), with \texttt{qwen2.5:32b} (0.543) and \texttt{gemma2:27b} (0.562) trailing behind, while \texttt{llama3.2} showed the lowest performance across both empathy and adherence.
Based on these considerations, we selected \texttt{deepseek-r1:32b} as the LLM-backbone for our prototype. This was mainly given by its higher \textbf{EmpathySupport}, which is aligned with the objectives of empathic prompting, and its overall stronger adherence to the system prompt. Despite \texttt{Gemma2} and \texttt{Qwen2.5} showing competitive but lower adherence scores, the qualitative review suggested that \texttt{DeepSeek}'s conversational style better balanced fluidity and affective alignment, making it the preferred choice for subsequent implementation.

Concerning system-level performance metrics, the obtained results are reported in Table~\ref{tab:model_perf}.

\begin{table}[!h]
\centering
\resizebox{\linewidth}{!}{
\begin{tabular}{lccc}
\toprule
\textbf{Model} & \textbf{Time (s)} & \textbf{Output Tokens} & \textbf{Tokens/s (TPS)} \\
\midrule
\texttt{\textbf{llama3.2:latest}} & \textbf{1.62 (1.27)} & \textbf{204.76 (184.39)} & \textbf{107.26 (32.16)} \\
\texttt{gemma2:27b}      & 6.70 (1.77) & 214.38 (58.11)  & 31.92 (0.57)  \\
\texttt{deepseek-r1:32b} & 23.31 (10.22) & 636.38 (281.63) & 27.22 (0.49)  \\
\texttt{\underline{qwen2.5:32b}}     & \underline{6.61 (3.00)} & \underline{174.19 (81.68)}  & \underline{26.09 (0.98)}  \\
\bottomrule
\end{tabular}
}
\caption{System-level performance metrics across backbone models. Values are reported as mean (std). Lowest values are bolded while second best results are undelined.}
\label{tab:model_perf}
\end{table}
 
The results highlight that \texttt{llama3.2:latest} achieved the fastest responses ($M = 1.62s$, $SD = 1.27$) and highest throughput ($M = 107.26$ TPS), but generated relatively short outputs ($M = 204.76$ tokens, $SD = 184.39$). \texttt{gemma2:27b} showed more verbose and stable outputs ($M = 214.38$ tokens, $SD = 58.11$) but with moderate latency ($M = 6.70s$). \texttt{qwen2.5:32b} performed similarly in speed ($M = 6.61s$) but produced shorter responses ($M = 174.19$ tokens).

In contrast, \texttt{deepseek-r1:32b} generated significantly longer and richer outputs ($M = 636.38$ tokens, $SD = 281.63$), more than tripling the verbosity of the other models, albeit with longer response times ($M = 23.31s$). \textcolor{black}{This is reasonable considering that this generation time includes time to stream tokens for the reasoning step (which is preliminary to the final user answer), which is only present in this model (among the selected ones)}. Despite lower throughput ($M = 27.22$ TPS), DeepSeek’s capacity to integrate contextual and empathic cues was superior, as confirmed by the LLM-as-a-Judge results.
It is worth observing that these tokens include both reasoning and an answer to the user. This provides an additional advantage for our logging system: having explainability for each considered answer.

\textcolor{black}{
These findings reinforce that DeepSeek is an appropriate backbone for our empathic prompting prototype: although slower, it provides strong instruction-following and reasoning-oriented behavior \cite{bi2024deepseek}, which is beneficial when responses must integrate multiple explicit constraints (dialogue history, safety rules, and affective cues). Moreover, external benchmarks on emotional-intelligence-style tasks (e.g., EQ-Bench) indicate that DeepSeek-r1 is one of the best open-weight LLMs in emotional reasoning~\cite{paech2023eq}. 
Despite this, DeepSeek is a text-only LLM and therefore cannot natively ingest or ground biometric signals from a stream as perceptual input; this is why in our implementation, we reduced the sensor stream to a compact affect tuple (Emotion, Intensity, Valence, Arousal) and then translates it into explicit natural-language instructions and constraints in the System Prompt, so that the affect signal becomes a readable and auditable conditioning variable for standard instruction following. This design choice is consistent with evidence that structured prompt formats can improve controllability and reduce ambiguity when the model is instructed on how to interpret fields and how to use them \cite{gao2025jsontuning,koshorek2025structured}.
Such choices aligns with the central aim of our work: to deliver a proof-of-concept that isolates and tests the effect of injecting robust, structured non-verbal affective context into the prompt, and to examine how this additional context modulates the coherence, tone, and dialogic flow of responses generated by a capable LLM.
}

\subsection{Pilot Usability Study}

This preliminary usability study was conducted exclusively with members of the research and development team. The goal was not to perform a formal empirical study, but rather to validate the technological functioning of the system and to gather initial impressions of its usability in a controlled and ethically safe context~\cite{nielsen1995applying,sohaib2010integrating}. 
\textcolor{black}{In line with established guidance on pilot studies, this staged internal evaluation was intentionally designed as a formative proof-of-concept to refine procedures and reduce methodological and ethical risk before external recruitment~\cite{thabane2010tutorial,eldridge2016defining}. Importantly, the internal participants did not contribute to system implementation; they were psychologists and researchers from the same lab for a controlled feasibility check. External validation with independent participants and raters is the next step following ethics approval.}
For the mentioned reasons, only internal staff, all fully aware of the nature and scope of the test, participated~\footnote{
\textcolor{black}{It is worth noticing that restricting participants may introduce bias and, for this reason, we treat the current results as formative and non-generalizable.}
}.
As colleagues directly involved in the project, participants were fully informed of the purpose of the sessions and the sensitivity of the topic, and their involvement was entirely voluntary.

The procedure followed an \textit{expert usability evaluation} approach. Each staff member interacted with the system and subsequently provided structured feedback on system performance, conversational flow, and perceived usability. By restricting the test to internal collaborators who were already familiar with the project, we ensured that participation did not involve any risk of misunderstanding or exposure and that the evaluation remained focused on technological validation.  
The rationale for this internal assessment was ethical: before involving external users, it was essential to verify that the system operated as intended and that its outputs could be reliably managed. This staged approach allows us to refine the prototype and address technical or design issues, ensuring that future empirical studies with end users will be based on a stable, well-tested system. 

\subsubsection{Participants}
The study involved a total of five participants, with ages ranging from 25 to 63 years (M = 40.4, SD = 19.3). The gender distribution was balanced, with three identifying as category 1 and two as category 2. In terms of education, participants had at least a high school diploma.
On a 7-point scale of technological familiarity, participants reported relatively high levels (M = 5.0, range = 3–6). Prior to the study, most participants (n = 4) had already interacted with chatbots, whereas one participant reported no such experience.


\subsubsection{Measurements}

To evaluate the user experience with the conversational agent, we employed a set of validated psychometric scales that target three core dimensions: system usability, perceived empathy, and specific qualities of human-agent interaction. All questionnaires were developed and administered in Italian through the Qualtrics online survey platform.
System usability was measured using the System Usability Scale (SUS; \cite{brooke1996sus}). This widely adopted 10-item instrument provides a holistic assessment of the effectiveness, efficiency, and subjective user satisfaction of a system.
To assess the agent's affective capabilities, we administered the Perceived Empathy of Technology Scale (PETS; \cite{schmidmaier2024perceived} ). This scale is particularly suited for affective computing contexts, measuring both the system's Emotional Responsiveness (PETS-ER) and its perceived Understanding and Trust (PETS-UT). To ease comprehension, we codified the entire scale as EMP in our results section.
Finally, to capture nuanced aspects of the interaction, we selectively adopted three subscales from the Godspeed Questionnaire Series (GQS; \cite{bartneck2008godspeed}, \cite{bartneck2023godspeed}), a standard instrument in Human-Agent Interaction research. Specifically, we administered the items for Likeability (LIKE), Perceived Intelligence (COMP), and Perceived Safety (SAFE). The Anthropomorphism and Animacy subscales were intentionally excluded, as their relevance is greater in human-robot interaction and less pertinent to the evaluation of a disembodied conversational AI.

\subsubsection{Experimental Procedure}

Our main goal for the experimental procedure amounts to the verification and validity of our Empathic Prompting system, i.e., the integration of textual data and facial emotional expressions parameters to support human–AI interaction.
\textcolor{black}{This evaluation is framed as a formative proof-of-concept: the study is designed to verify end-to-end feasibility and gather initial usability impressions, and all reported metrics should be interpreted as descriptive and preliminary.
\\
It is also worth noting that we did not include a dedicated text-only baseline condition (i.e., the same chatbot without affective descriptors) because the primary objective was to validate the complete sensing-to-prompt pipeline under strict local-deployment and privacy constraints, and to iterate on the interaction protocol before running comparative efficacy tests. Moreover, previous work has demonstrated that injecting emotionally grounded information improves performance compared to classical approaches~\cite{arjmand2024empathic}.
}

Given the sensitive nature of the data collected (including emotional responses and conversational content) all participants were fully informed about the study’s aims and procedures. Participants were affiliated with 
University of xxx~\footnote{Anonymized due to double blind peer review process.},
ensuring a high level of awareness and ethical compliance throughout the experimental process.
The procedure consists of these steps:
\begin{enumerate}
    \item \textbf{Participant Onboarding}: Upon arrival, participants were seated comfortably and given a concise verbal overview of the study. The researcher introduced the experiment with the following explanation: ``\textit{We are conducting a study on Artificial Intelligence, exploring how textual data can be integrated with facial emotional expressions. After signing the informed consent form, you will view a series of images and then engage in a 5-minute conversation with a chatbot about what you saw and how you felt.}''
    \item \textbf{Informed Consent}: Participants were then presented with the informed consent questionnaire, which detailed the study’s purpose, procedures, risks, and data protection measures. Each participant was assigned a unique identification code, which was verified to ensure consistency across all data collection phases. As all participants were researchers, they were fully aware of the ethical implications and voluntarily agreed to participate.
    \item \textbf{Visual Priming Stimulation}: Upon providing consent, participants underwent a \textcolor{black}{validated visual priming procedure~\cite{Kurdi2017OASIS}}. They viewed a sequence of images structured to elicit discrete primary emotions by systematically interspersing emotionally evocative stimuli with neutral ones, a design intended to mitigate affective habituation. The presentation duration for each stimulus was fixed at one second.
    \item \textbf{Chatbot Interaction} Before initiating the conversation, participants received the following instruction: ``\textit{Begin a conversation with the chatbot by answering the question $<$ Talk about what you saw and what emotions you felt regarding the images you just viewed $>$.}''. Then, participants engaged in a free-form dialogue with the chatbot for a duration of five minutes. This interaction was intended to capture spontaneous emotional and cognitive reflections in response to the visual stimuli.
    \item \textbf{Post-Interaction Questionnaire}: Immediately after the chatbot session, participants were directed to complete a final questionnaire designed to assess their impressions of the chatbot interaction and their emotional state. The identification code used in the initial consent form was re-entered to ensure data linkage and integrity.
\end{enumerate}

\subsubsection{Quantitative Results}

\paragraph{Reliability analysis}

\textcolor{black}{
\begin{table}[!h]
\centering
\caption{Cronbach's $\alpha$ reliability results with 95\% confidence intervals.}
\begin{tabular}{lcc}
\hline
\textbf{Scale} & \textbf{Cronbach's $\alpha$} & \textbf{95\% CI} \\
\hline
SUS & 0.647  & [-0.119, 0.958]\\
EMP      & 0.727 & [0.136, 0.968] \\
LIKE     & 0.771 & [0.146, 0.973] \\
COMP     & 0.789 & [0.215, 0.976] \\
SAFE     & 0.477 & [-1.641, 0.942] \\
\hline
\end{tabular}
\label{tab:cronbach}
\end{table}
}

\textcolor{black}{The analysis of internal reliability shows that the \textbf{COMP} and \textbf{LIKE} scales achieved the highest levels of internal consistency (Cronbach’s $\alpha > 0.77$), indicating good coherence among their items. The \textbf{EMP} scale also reported an acceptable level of reliability ($\alpha = 0.727$), while \textbf{SUS} fell within a moderate range ($\alpha = 0.647$)~\footnote{The Negative SUS components were reverse coded for coherency in the Cronbach's $\alpha$ calculation.}}. This is, however, worth analyzing, considering that the contradiction in different positive/negative constructs may provide insights for future development of our system.
Similarly, the \textbf{SAFE} scale showed low reliability ($\alpha = 0.477$) with a very wide confidence interval- It is important to note that these results should be interpreted with caution, as the limited number of participants and the presence of only three questions in the \textbf{SAFE} construct reduce statistical stability (also considering the kind of questions included in the questionnaire). \textcolor{black}{Despite these limitations, the analysis remains valuable for identifying critical areas that require refinement in future iterations of the instrument and provides preliminary indications of the psychometric robustness of the adopted scales. This is also valid considering that the sample size ($N=5$) could provide unstable estimates~\cite{taber2018use}. For this reasons, we reported such results for transparency, even if they cannot be a strong evidence of validity~\cite{sauro2011designing}.
}


\paragraph{Questionnaire Analysis}

In this Section, we report the descriptive results of the scores obtained in all the considered scales.

\begin{figure}[!h]
    \centering
    \begin{subfigure}[t]{0.48\linewidth}
        \centering
        \includegraphics[width=\linewidth]{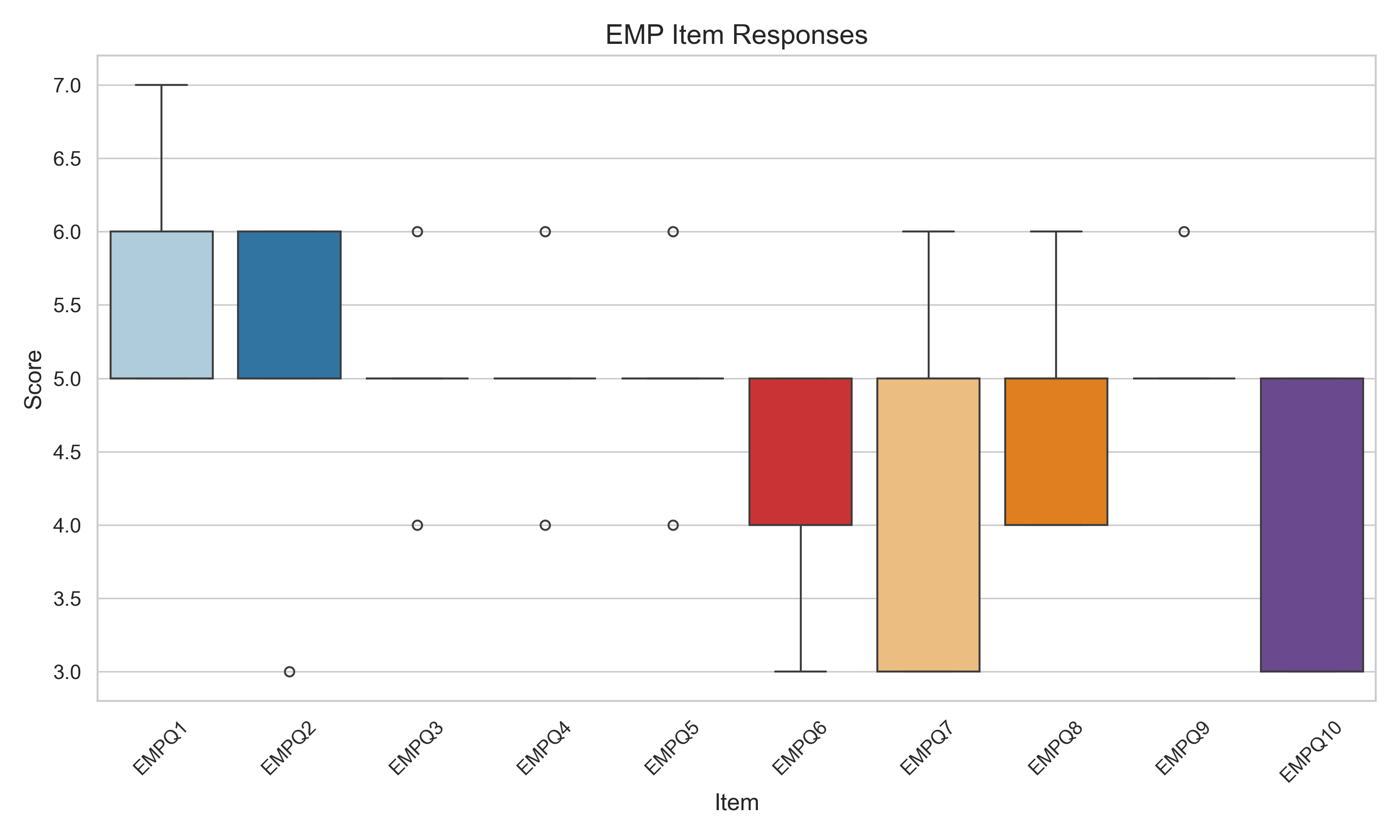}
        \caption{Distribution of item responses for the EMP construct.}
        \label{fig:emp_boxplot}
    \end{subfigure}
    \hfill
    \begin{subfigure}[t]{0.48\linewidth}
        \centering
        \includegraphics[width=\linewidth]{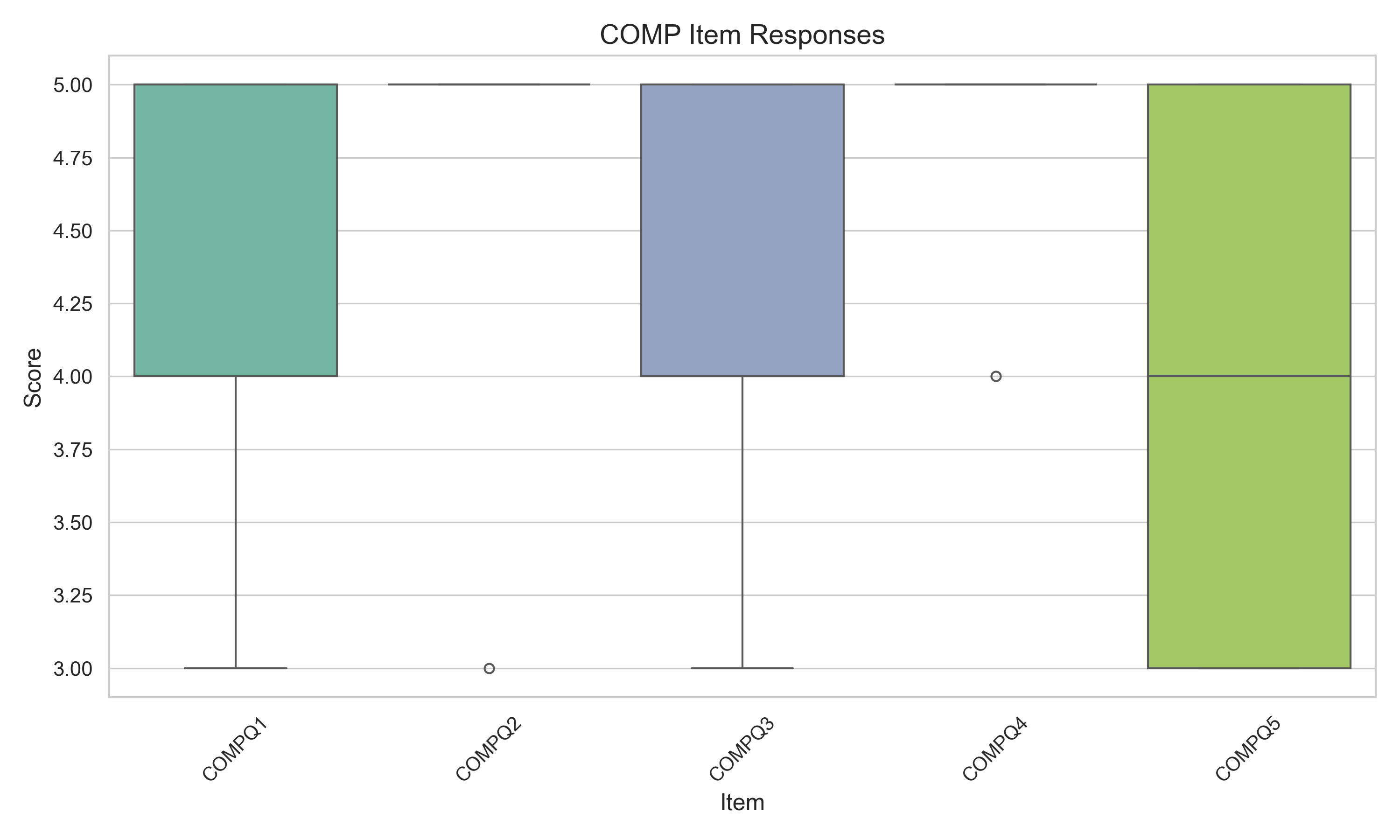}
        \caption{Distribution of item responses for the Comprehension (COMP) construct.}
        \label{fig:comp_boxplot}
    \end{subfigure}
    \caption{Boxplots illustrating participant responses across two constructs of the questionnaire: EMP and COMP.}
    \label{fig:emp_comp_boxplots}
\end{figure}

Figure~\ref{fig:emp_comp_boxplots} illustrates the distribution of responses across the EMP and COMP constructs.  

For the EMP construct (Figure~\ref{fig:emp_boxplot}), responses were more heterogeneous across items. Higher scores were observed for statements capturing emotional sensitivity and social presence, such as “the system considered my mental state” (EMPQ1), “the system seemed emotionally intelligent” (EMPQ2), “the system showed interest in me” (EMPQ5), and “I trusted the system” (EMPQ9). By contrast, items that probed deeper empathic processes, such as “the system helped me manage an emotional situation” (EMPQ6), “the system understood my goals” (EMPQ7), and “the system understood my intentions” (EMPQ10), received lower and more dispersed ratings. This suggests that while the system was perceived as attentive and affectively aware, it was less consistently judged as capable of providing instrumental support or goal-oriented empathy. The variability in responses is likely influenced by the conversational task design, which did not involve explicit objectives or emotionally demanding scenarios, thereby limiting opportunities for participants to evaluate these dimensions.  

In contrast, the COMP construct (Figure~\ref{fig:comp_boxplot}) shows uniformly higher ratings across all items, with limited variability. Participants consistently evaluated the system as competent, intelligent, and judicious, as reflected in high scores for items such as “the system was intelligent” (COMPQ4) and “the system was judicious” (COMPQ5). Importantly, no item in this construct elicited systematically negative evaluations. 

\begin{figure}[!h]
    \centering
    \begin{subfigure}[t]{0.48\linewidth}
        \centering
        \includegraphics[width=\linewidth]{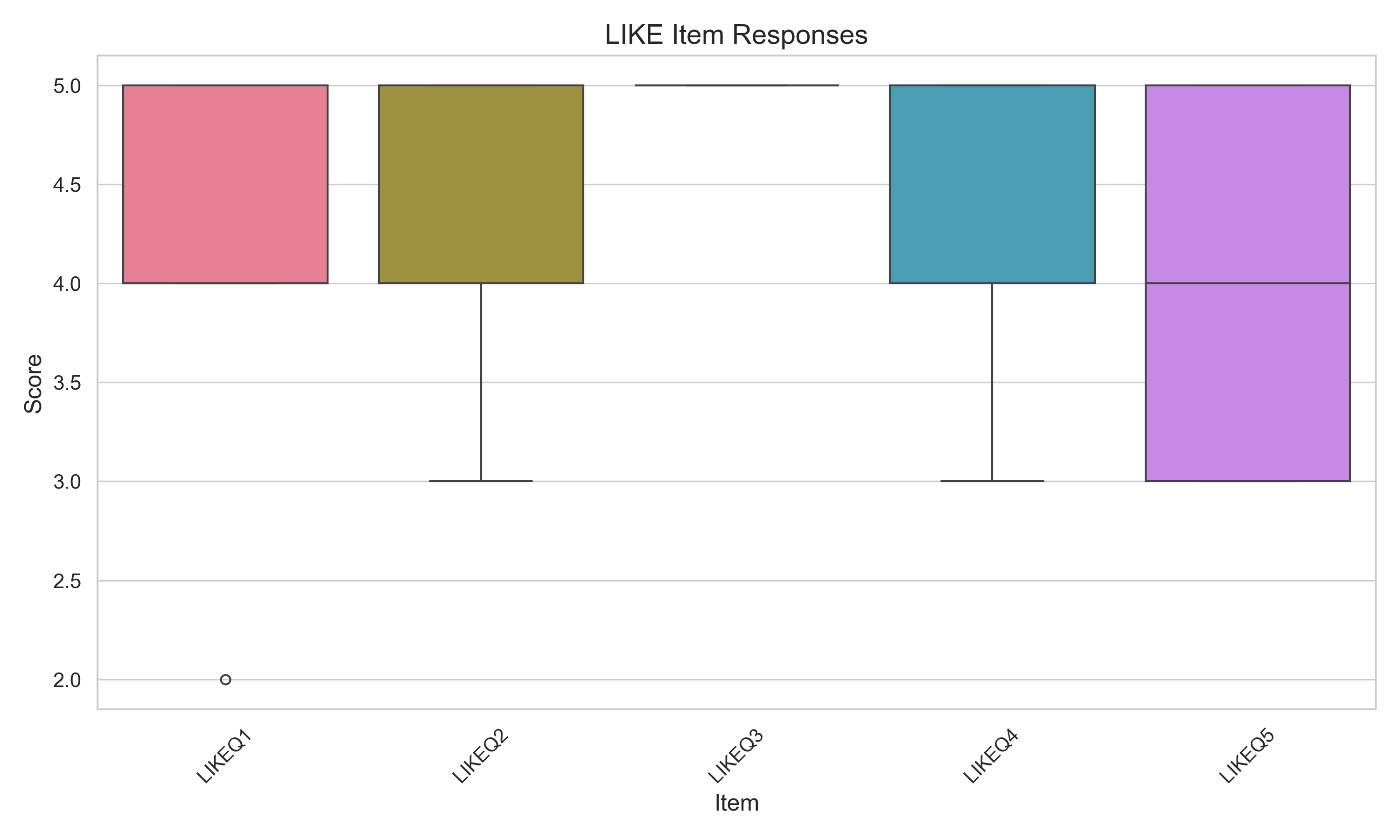}
        \caption{Distribution of item responses for the LIKE construct.}
        \label{fig:like_boxplot}
    \end{subfigure}
    \hfill
    \begin{subfigure}[t]{0.48\linewidth}
        \centering
        \includegraphics[width=\linewidth]{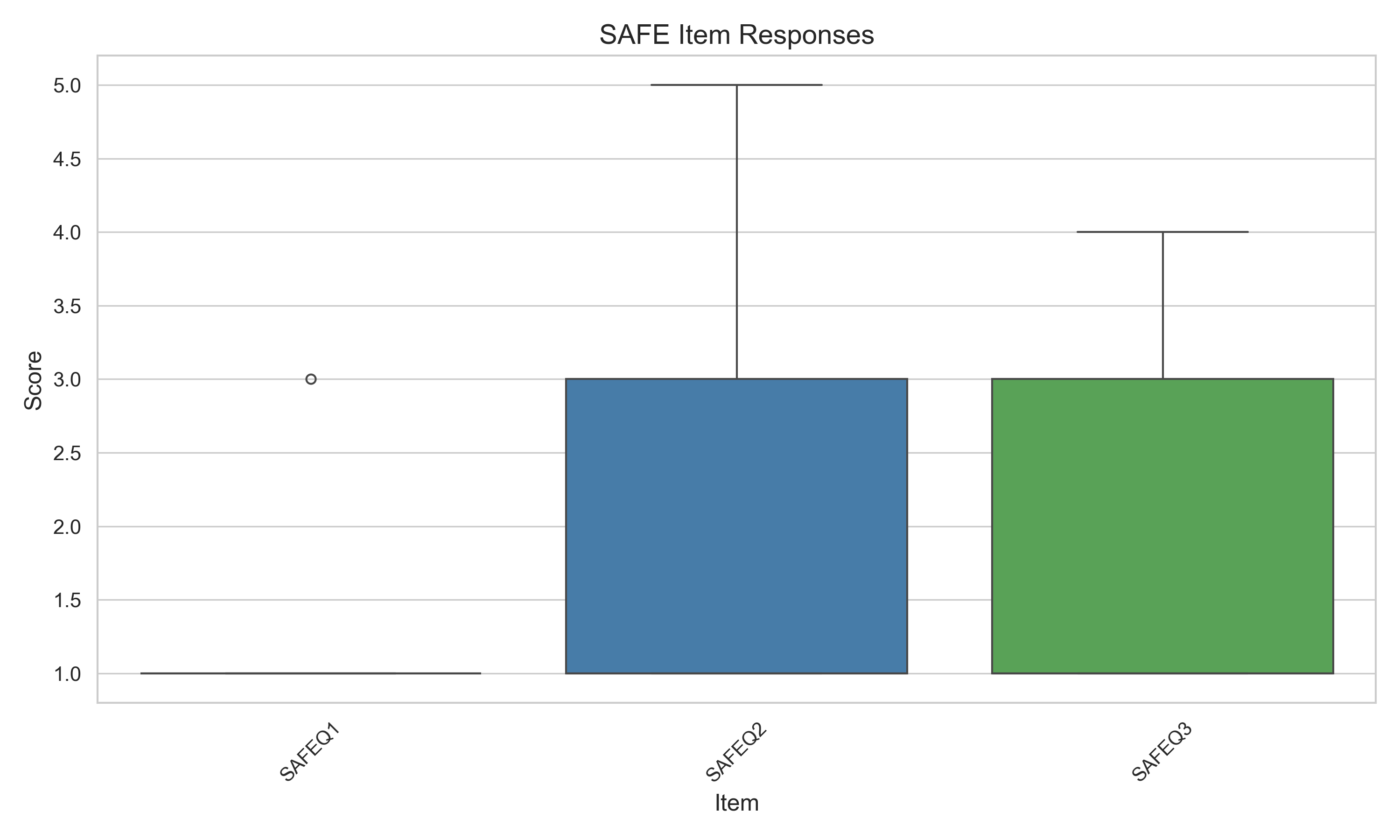}
        \caption{Distribution of item responses for the SAFE construct.}
        \label{fig:safe_boxplot}
    \end{subfigure}
    \caption{Boxplots illustrating participant responses across two constructs of the questionnaire: LIKE and SAFE.}
    \label{fig:like_safe_boxplots}
\end{figure}

Figure~\ref{fig:like_safe_boxplots} illustrates participant responses to the LIKE and SAFE constructs.  
For the LIKE construct (Figure~\ref{fig:like_boxplot}), responses show generally high ratings across all items, indicating that the chatbot was perceived as pleasant, friendly, and agreeable. Items such as “the system was friendly” (LIKEQ2) and “the system was amiable” (LIKEQ3) consistently scored highly, suggesting that the system successfully conveyed a socially positive persona. Some variability was observed for the item “the system was beautiful” (LIKEQ5), which received more dispersed ratings. This outcome is unsurprising, as aesthetic judgments are less meaningful in the context of human–computer interaction with a text-based agent, and therefore more subject to individual interpretation. Overall, the LIKE construct highlights that participants attributed to the system qualities of warmth and social acceptability, reinforcing its perceived usability in affective contexts.  

In contrast, the SAFE construct (Figure~\ref{fig:safe_boxplot}) yielded lower and more variable responses. Items such as “I felt calm rather than agitated” (SAFEQ2) and “I felt serene rather than surprised” (SAFEQ3) tended toward mid-range scores, while “I felt relaxed rather than anxious” (SAFEQ1) displayed particularly low values and higher dispersion. This pattern suggests that participants were less confident in attributing a sense of safety or emotional containment to the system. The reduced reliability of the SAFE scale (as also reflected in its Cronbach’s $\alpha$) can partly be explained by the limited number of items (only three) and their conceptual heterogeneity, which makes them less cohesive as a construct. Nonetheless, the results provide useful preliminary insights, indicating that while the chatbot is consistently experienced as likable and socially pleasant, its capacity to instill feelings of safety and emotional reassurance remains weaker and more context-dependent.


\begin{figure}[!h]
    \centering
    \begin{subfigure}[t]{0.48\linewidth}
        \centering
        \includegraphics[width=\linewidth]{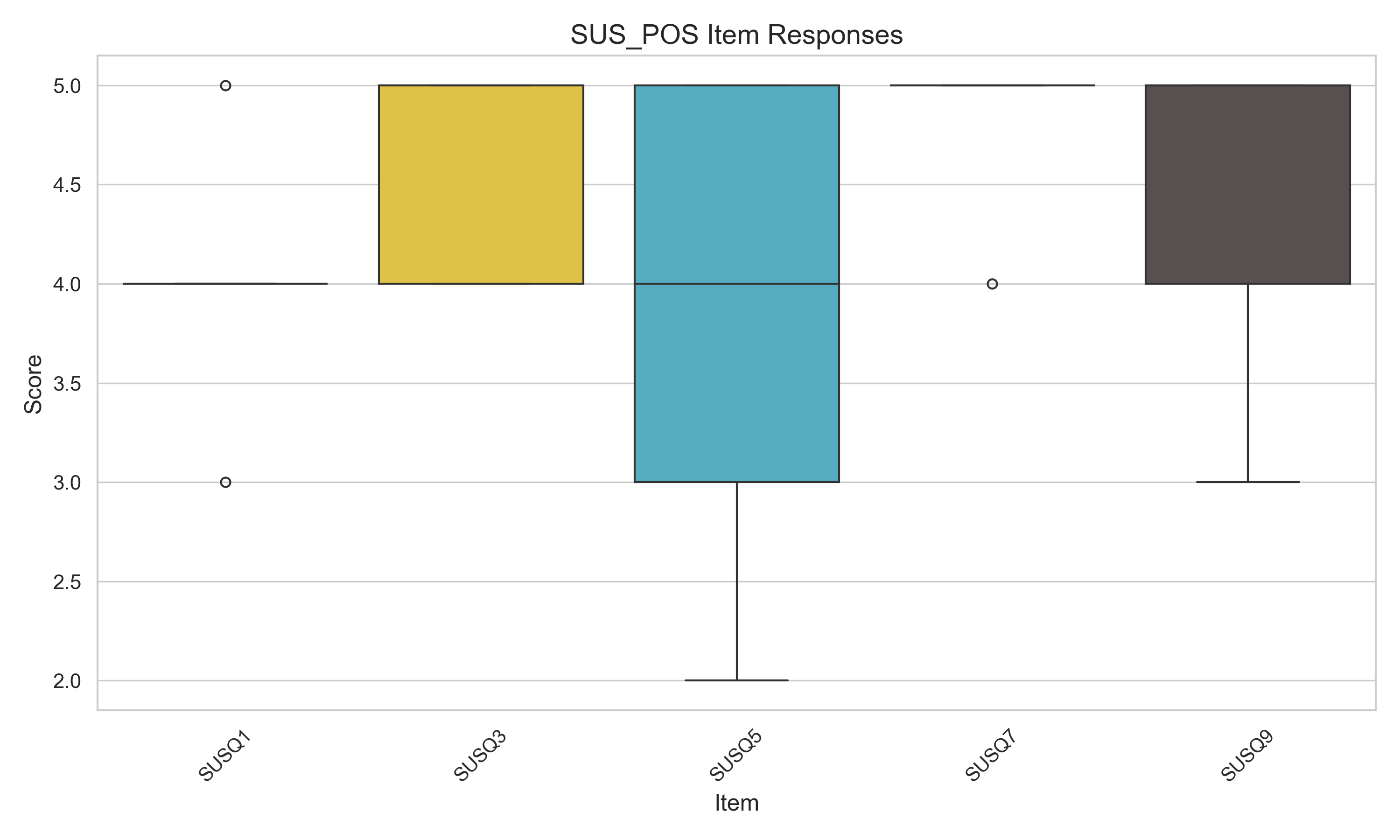}
        \caption{Distribution of item responses for the positive SUS items.}
        \label{fig:sus_pos_boxplot}
    \end{subfigure}
    \hfill
    \begin{subfigure}[t]{0.48\linewidth}
        \centering
        \includegraphics[width=\linewidth]{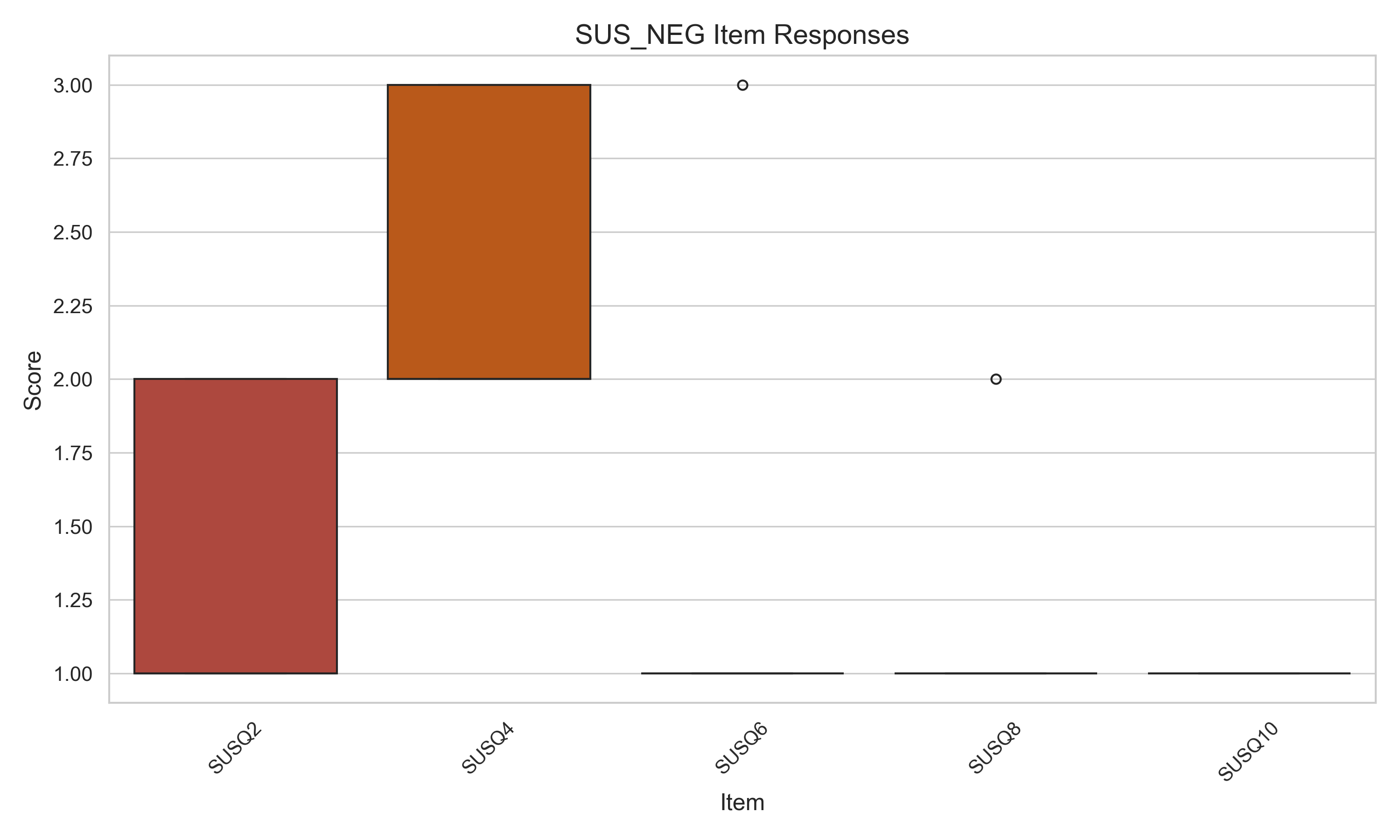}
        \caption{Distribution of item responses for the negative SUS items.}
        \label{fig:sus_neg_boxplot}
    \end{subfigure}
    \caption{Boxplots illustrating the spread of responses across the System Usability Scale (SUS), separated into positive and negative items.}
    \label{fig:sus_boxplots}
\end{figure}

Figure~\ref{fig:sus_boxplots} reports instead the distribution of responses across the System Usability Scale (SUS), separated into positive (Figure~\ref{fig:sus_pos_boxplot}) and negative (Figure~\ref{fig:sus_neg_boxplot}) items.  

For the positive items, responses clustered around relatively high values, indicating that participants generally perceived the system as easy to use, well integrated, and confidence-inducing. Items such as “I found the system simple to use” (SUSQ3), “I felt confident using the system” (SUSQ9), and “I found the system’s functions well integrated” (SUSQ5) achieved consistently strong ratings. These results suggest that, despite its experimental nature, the prototype already conveyed a sense of usability and coherence that is typical of more mature systems.  

Conversely, the negative items received very low scores, which is an encouraging signal in SUS interpretation. Low agreement with statements such as “I found the system unnecessarily complex” (SUSQ2), “I needed the support of a technical person to use the system” (SUSQ4), or “I found inconsistencies in the system” (SUSQ6) suggests that participants did not experience substantial barriers or frustrations during interaction. However, some dispersion was observed, especially for SUSQ4, indicating that a subset of users felt they might still require support to fully exploit the system. 

\subsection{Qualitative System Analysis}

While our quantitative analysis (e.g., questionnaire responses and correlation analyses) doesn't show one of the main advantages of our Empathic Prompting system: analyzing how the user emotionally reacts when the LLM answers one of her/his messages. 
To carry out this analysis, we conducted a qualitative analysis of the emotional flow with one member of the team. This member of the team never used/or implemented our prototype before, but it was aware of how the entire system works.
The goal of this analysis was to examine how the empathic prompting system handled affective input over time and whether the integration of non-verbal cues with LLM responses produced coherent conversational outcomes.

\begin{figure}[!h]
    \centering
    \includegraphics[width=\linewidth]{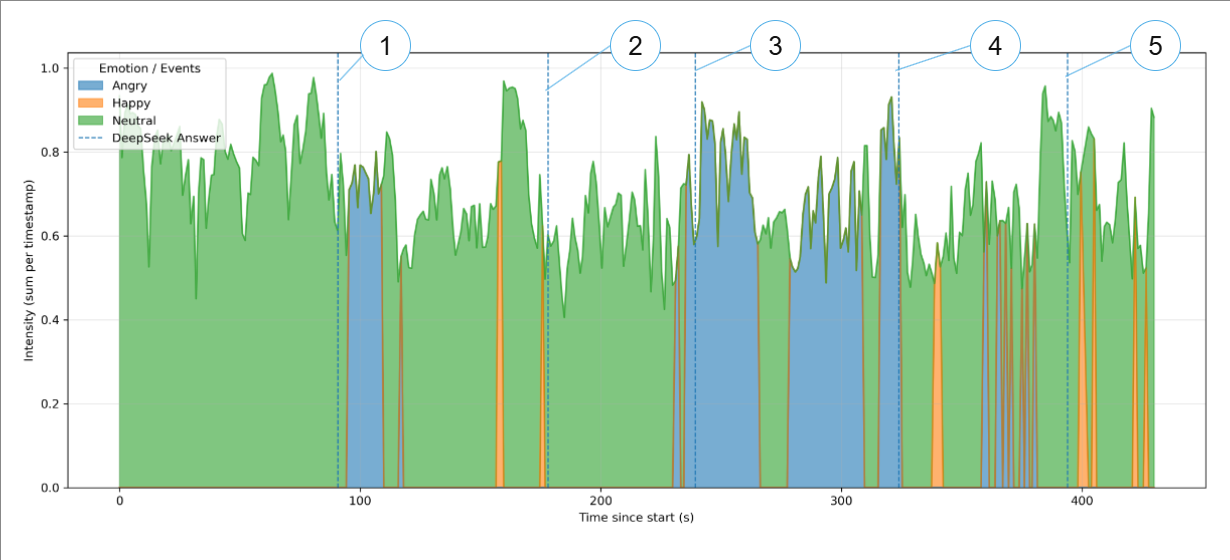}
    \caption{Temporal distribution of detected emotions (Neutral, Happy, Angry) during the interaction. Vertical dashed lines (1–5) indicate messages generated by DeepSeek in response to user prompts, which are qualitatively analyzed in the study.}
    \label{fig:temporal_distribution}
\end{figure}

Figure~\ref{fig:temporal_distribution} shows the temporal distribution of emotional signals (Neutral, Happy, Angry) captured during an interaction session, overlaid with vertical dashed markers corresponding to the system’s responses (DeepSeek outputs) following user prompts. This representation allows us to align biometric affective fluctuations with conversational events, enabling a more fine-grained inspection of how the system reacted to shifts in user emotion. By qualitatively examining these episodes, we can evaluate whether empathic prompting produced answers that were timely, contextually appropriate, and aligned with the emotional state of the participants, thereby testing the functional coherence of the full architecture.

\textcolor{black}{A granular analysis of interaction logs revealed that inconsistencies between facial expression and textual sentiment often reflect distinct behavioral states rather than simple misalignment. We identified two primary patterns successfully navigated by the Empathic Prompting architecture.
\\
By analyzing the interaction logs, we identified distinct patterns of incongruence that the system successfully navigated.
\\
First, we observed a phenomenon of Cognitive Masking \cite{cacioppo1985semantic} in the initial phase (Turn 1). Here, the user provided highly emotional self-disclosure, yet the biometric sensor registered a "Neutral" state (Intensity: 0.96), likely due to the cognitive load of text formulation suppressing spontaneous affect. The system, adhering to prompt instructions, validated the text while acknowledging the "composed" demeanor. This intervention appeared to act as a catalyst: in the subsequent turn (Turn 2), the user's expression shifted markedly to "Happiness" (Intensity: 0.78), suggesting the system established a resonant feedback loop that authorized the release of suppressed emotion.
\\
Second, a critical incongruence emerged in Turn 4, where the user expressed "tenderness" ("Mi commuove") while the FaceReader detected "Anger" (Intensity: 0.65). Psychophysiologically, this likely reflects corrugator supercilii activation \cite{balconi2013emotional} associated with deep empathy, often misclassified as aggression by automated tools. Instead of hallucinating hostility, the system integrated the semantic context with the biometric input, successfully re-contextualizing the "Anger" signal as "underlying emotional tension".
\\
From this analysis, we observed the system' capability of adhering to the empathic prompting guidelines, producing contextually appropriate and affectively responsive interactions that managed these multimodal frictions effectively. Nonetheless, limitations were observed, particularly hallucinations related to physical activity and some textual redundancies, which may hinder conversational fluency.
}


\section{Discussion and Conclusions}
\label{sec:disc_concl}

We here introduced \emph{Empathic Prompting}, a novel framework designed to integrate non-verbal affective cues into LLM conversations. By combining biometric input from a professional Face Reader with contextualized prompting, the system was able to align generated responses with the user’s emotional state.

The empirical results from our internal usability study showed that, across multiple constructs, participants consistently rated the system as usable and coherent, with particularly high evaluations in the domains of empathy and perceived intelligence.


In particular, with respect to \textbf{RQ1} (“Does integrating non-verbal context through prompting improve perceived empathy and safety-aligned behavior?”), Our internal usability study showed that participants consistently rated the system as usable and coherent, with particularly high evaluations in the domains of empathy and perceived intelligence. These findings suggest that non-verbal integration improves the perception of empathy and contributes to a general sense of emotional intelligence and cognitive reliability. Considering instead safety, users were uncertain about how safe this conversational methodology can be.

In relation to \textbf{RQ2} (“How do users experience smoothness and alignment when generation is conditioned on real-time affect?”), Our qualitative analysis of emotional flow highlighted that the system was able to track and adapt to user affective shifts over time. Chatbot responses were broadly consistent with emotional cues and produced interactions that participants described as fluid and contextually aligned. Although some inconsistencies were observed, the analysis confirms that conditioning generation on affective signals can enhance conversational smoothness and emotional alignment, \textcolor{black}{maintaining a coherent and contextually appropriate exchange}.
Finally, usability constructs exhibited low and uncertain scores on negative aspects, which is a positive signal of robustness towards RQ2.


It is worth mentioning that, considering the ethical and methodological perspective, we deliberately adopted a two-step approach. The first step focused on system design, development, and performance validation, complemented by a minimal internal usability test to verify whether users perceived the interface as working. After this technological validation, we now proceed with the second step, which consists of a full empirical trial with external participants, which will require prior approval from the ethics committee.
This two-phase approach is a strength of our work, as it ensures that ethically sensitive testing in real-world contexts is preceded by a robust and safe system design phase.

However, some limitations are worth mentioning.  The dataset adopted for our preliminary LLM evaluation consisted of representative conversations validated by a psychologist, but its size remains limited.
The usability test was carried out on a small number of participants, limiting statistical generalization. \textcolor{black}{In the test, we also did not measure emotion recognition accuracy (considering the adopted Face Reader as an oracle).}
Moreover, the construct related to perceived safety outlined general low judgments, requiring refinement and task-specific investigation.
\textcolor{black}{Finally, considering a single point estimate to represent affect over an interaction episode may be limiting: employing a compact distribution representation (e.g., emotion histograms) may provide more informative and stable summaries than a single label. This may be particularly effective in our use case, since they could preserve brief deviations from neutrality during reading/typing while remaining auditable and easy to map into prompt-level constraints.
}
Considering these limitations, future work will include larger user studies \textcolor{black}{(including objective emotions evaluations)}, novel development to refine the exploitation of affective signals, and multiple use cases and datasets to evaluate our approach in domains such as healthcare, education, and support services.

To conclude, this preliminary study provides initial evidence that \emph{Empathic Prompting} can improve the quality of human–AI interaction by embedding affective awareness into language generation.  It is not yet clear how conversations enriched with implicit affective information will change human–AI interactions. While our findings demonstrate the feasibility and potential of emotional-augmented prompting, its long-term implications remain to be fully understood.

\section*{Funding details}
No funding was received for conducting this study.

\section*{Disclosure statement}
The authors report there are no competing interests to declare.

\section*{Data availability statement}
The data supporting the findings of this study are available from the corresponding author upon reasonable request.  
Due to the involvement of biometric and emotional information, data are not publicly available to preserve participants’ privacy and comply with ethical restrictions.

\section*{Data deposition}
No datasets were deposited in public repositories. The study relies on internally collected usability and evaluation data from a small-sample pilot study (N=5) conducted within the research group environment.


\section*{Ethical Declaration}
This study was reviewed and approved by the Ethical Committee of the University of XXXX~\footnote{Anonym due to double blind.} (Ethical Committee for Research, Protocol n.\ 117651, 17~September~2025). 
The Committee determined that the research complied with institutional and national ethical standards and did not involve the collection of sensitive, biometric, or behavioral data. 
All participants provided informed consent prior to participation, and all procedures were conducted in accordance with the Declaration of Helsinki and the European GDPR framework for data protection.

\section*{Acknowledgments}
ChatGPT was utilized to refine, rephrase, and improve the clarity of the English text. The authors reviewed, corrected, and approved all final content.

\bibliographystyle{elsarticle-num-names} 

\bibliography{bibliography_new}

@String{Computing = "Computing" }

@String{Computer = "{IEEE} Computer" }

@String{Academic = "Academic Press" }

@String{Springer = "Springer-Verlag" }

@article{Khot2022Decomposed,
  title={Decomposed prompting: A modular approach for solving complex tasks},
  author={Khot, Tushar and Trivedi, Harsh and Finlayson, Matthew and Fu, Yao and Richardson, Kyle and Clark, Peter and Sabharwal, Ashish},
  journal={arXiv preprint arXiv:2210.02406},
  year={2022}
}

@article{ma2025context,
  title={In-Context Examples Matter: Improving Emotion Recognition in Conversation with Instruction Tuning},
  author={Ma, Hui and Zhang, Bo and Hu, Jinpeng and Shi, Zenglin},
  journal={arXiv preprint arXiv:2508.11889},
  year={2025}
}

@article{Wei2022Emergent,
  title={Emergent abilities of large language models},
  author={Wei, Jason and Tay, Yi and Bommasani, Rishi and Raffel, Colin and Zoph, Barret and Borgeaud, Sebastian and Yogatama, Dani and Bosma, Maarten and Zhou, Denny and Metzler, Donald and others},
  journal={arXiv preprint arXiv:2206.07682},
  year={2022}
}

@article{priyadarshana2024prompt,
  title={Prompt engineering for digital mental health: a short review},
  author={Priyadarshana, YHPP and Senanayake, Ashala and Liang, Zilu and Piumarta, Ian},
  journal={Frontiers in Digital Health},
  volume={6},
  pages={1410947},
  year={2024},
  publisher={Frontiers Media SA}
}

@inproceedings{arjmand2024empathic,
  title={Empathic Grounding: Explorations using Multimodal Interaction and Large Language Models with Conversational Agents},
  author={Arjmand, Mehdi and Nouraei, Farnaz and Steenstra, Ian and Bickmore, Timothy},
  booktitle={Proceedings of the 24th ACM International Conference on Intelligent Virtual Agents},
  pages={1--10},
  year={2024}
}

@inproceedings{liu2025proactive,
  title={Proactive conversational agents with inner thoughts},
  author={Liu, Xingyu Bruce and Fang, Shitao and Shi, Weiyan and Wu, Chien-Sheng and Igarashi, Takeo and Chen, Xiang'Anthony'},
  booktitle={Proceedings of the 2025 CHI Conference on Human Factors in Computing Systems},
  pages={1--19},
  year={2025}
}

@article{schulhoff2024prompt,
  title={The prompt report: a systematic survey of prompt engineering techniques},
  author={Schulhoff, Sander and Ilie, Michael and Balepur, Nishant and Kahadze, Konstantine and Liu, Amanda and Si, Chenglei and Li, Yinheng and Gupta, Aayush and Han, HyoJung and Schulhoff, Sevien and others},
  journal={arXiv preprint arXiv:2406.06608},
  year={2024}
}

@article{SurveyPrompting2024,
  title={A systematic survey of prompt engineering in large language models: Techniques and applications},
  author={Sahoo, Pranab and Singh, Ayush Kumar and Saha, Sriparna and Jain, Vinija and Mondal, Samrat and Chadha, Aman},
  journal={arXiv preprint arXiv:2402.07927},
  year={2024}
}

@article{sorin2024large,
  title={Large language models and empathy: systematic review},
  author={Sorin, Vera and Brin, Dana and Barash, Yiftach and Konen, Eli and Charney, Alexander and Nadkarni, Girish and Klang, Eyal},
  journal={Journal of medical Internet research},
  volume={26},
  pages={e52597},
  year={2024},
  publisher={JMIR Publications Toronto, Canada}
}

@article{gu2024survey,
  title={A survey on llm-as-a-judge},
  author={Gu, Jiawei and Jiang, Xuhui and Shi, Zhichao and Tan, Hexiang and Zhai, Xuehao and Xu, Chengjin and Li, Wei and Shen, Yinghan and Ma, Shengjie and Liu, Honghao and others},
  journal={arXiv preprint arXiv:2411.15594},
  year={2024}
}

@article{grattafiori2024llama,
  title={The llama 3 herd of models},
  author={Grattafiori, Aaron and Dubey, Abhimanyu and Jauhri, Abhinav and Pandey, Abhinav and Kadian, Abhishek and Al-Dahle, Ahmad and Letman, Aiesha and Mathur, Akhil and Schelten, Alan and Vaughan, Alex and others},
  journal={arXiv preprint arXiv:2407.21783},
  year={2024}
}

@article{zheng2023judging,
  title={Judging llm-as-a-judge with mt-bench and chatbot arena},
  author={Zheng, Lianmin and Chiang, Wei-Lin and Sheng, Ying and Zhuang, Siyuan and Wu, Zhanghao and Zhuang, Yonghao and Lin, Zi and Li, Zhuohan and Li, Dacheng and Xing, Eric and others},
  journal={Advances in neural information processing systems},
  volume={36},
  pages={46595--46623},
  year={2023}
}

@article{zhu2023judgelm,
  title={Judgelm: Fine-tuned large language models are scalable judges},
  author={Zhu, Lianghui and Wang, Xinggang and Wang, Xinlong},
  journal={arXiv preprint arXiv:2310.17631},
  year={2023}
}

@article{skiendziel2019assessing,
  title={Assessing the convergent validity between the automated emotion recognition software Noldus FaceReader 7 and Facial Action Coding System Scoring},
  author={Skiendziel, Tanja and R{\"o}sch, Andreas G and Schultheiss, Oliver C},
  journal={PloS one},
  volume={14},
  number={10},
  pages={e0223905},
  year={2019},
  publisher={Public Library of Science San Francisco, CA USA}
}

@article{li2023cross,
  title={The cross-race effect in automatic facial expression recognition violates measurement invariance},
  author={Li, Yen-Ting and Yeh, Su-Ling and Huang, Tsung-Ren},
  journal={Frontiers in Psychology},
  volume={14},
  pages={1201145},
  year={2023},
  publisher={Frontiers Media SA}
}

@article{lewinski2014automated,
  title={Automated facial coding: validation of basic emotions and FACS AUs in FaceReader.},
  author={Lewinski, Peter and Den Uyl, Tim M and Butler, Crystal},
  journal={Journal of neuroscience, psychology, and economics},
  volume={7},
  number={4},
  pages={227},
  year={2014},
  publisher={Educational Publishing Foundation}
}

@article{koshorek2025structured,
  title={Structured RAG for Answering Aggregative Questions},
  author={Koshorek, Omri and Granot, Niv and Alloni, Aviv and Admati, Shahar and Hendel, Roee and Weiss, Ido and Arazi, Alan and Cohen, Shay-Nitzan and Belinkov, Yonatan},
  journal={arXiv preprint arXiv:2511.08505},
  year={2025}
}

@article{balconi2013emotional,
  title={Emotional contagion and trait empathy in prosocial behavior in young people: the contribution of autonomic (facial feedback) and balanced emotional empathy scale (BEES) measures},
  author={Balconi, Michela and Canavesio, Ylenia},
  journal={Journal of Clinical and Experimental Neuropsychology},
  volume={35},
  number={1},
  pages={41--48},
  year={2013},
  publisher={Taylor \& Francis}
}

@article{cacioppo1985semantic,
  title={Semantic, evaluative, and self-referent processing: Memory, cognitive effort, and somatovisceral activity},
  author={Cacioppo, John T and Petty, Richard E and Morris, Katherine J},
  journal={Psychophysiology},
  volume={22},
  number={4},
  pages={371--384},
  year={1985},
  publisher={Wiley Online Library}
}

@inproceedings{gao2025jsontuning,
    title = "{J}son{T}uning: Towards Generalizable, Robust, and Controllable Instruction Tuning",
    author = "Gao, Chang  and
      Zhang, Wenxuan  and
      Chen, Guizhen  and
      Lam, Wai",
    editor = "Che, Wanxiang  and
      Nabende, Joyce  and
      Shutova, Ekaterina  and
      Pilehvar, Mohammad Taher",
    booktitle = "Findings of the Association for Computational Linguistics: ACL 2025",
    month = jul,
    year = "2025",
    address = "Vienna, Austria",
    publisher = "Association for Computational Linguistics",
    url = "https://aclanthology.org/2025.findings-acl.1232/",
    doi = "10.18653/v1/2025.findings-acl.1232",
    pages = "24029--24055",
    ISBN = "979-8-89176-256-5"
}

@inproceedings{plantak2021usability,
  title={Usability and user experience of a chat application with integrated educational chatbot functionalities},
  author={Plantak Vukovac, Dijana and Horvat, Ana and {\v{C}}i{\v{z}}me{\v{s}}ija, Antonela},
  booktitle={International Conference on Human-Computer Interaction},
  pages={216--229},
  year={2021},
  organization={Springer}
}

@inproceedings{prpa2024challenges,
  title={Challenges and opportunities of LLM-based synthetic personae and data in HCI},
  author={Prpa, Mirjana and Troiano, Giovanni and Yao, Bingsheng and Li, Toby Jia-Jun and Wang, Dakuo and Gu, Hansu},
  booktitle={Companion Publication of the 2024 Conference on Computer-Supported Cooperative Work and Social Computing},
  pages={716--719},
  year={2024}
}

@inproceedings{marquardt2025ragatar,
  title={RAGatar: Enhancing LLM-driven Avatars with RAG for Knowledge-Adaptive Conversations in Virtual Reality},
  author={Marquardt, Alexander and Golchinfar, David and Vaziri, Daryoush},
  booktitle={2025 IEEE Conference on Virtual Reality and 3D User Interfaces Abstracts and Workshops (VRW)},
  pages={1604--1605},
  year={2025},
  organization={IEEE}
}

@misc{Ip_deepeval_2025,
author = {Ip, Jeffrey and Vongthongsri, Kritin},
license = {Apache-2.0},
month = aug,
title = {{deepeval}},
url = {https://github.com/confident-ai/deepeval},
version = {3.4.3},
year = {2025}
}

@article{agarwal2025gpt,
  title={gpt-oss-120b \& gpt-oss-20b Model Card},
  author={Agarwal, Sandhini and Ahmad, Lama and Ai, Jason and Altman, Sam and Applebaum, Andy and Arbus, Edwin and Arora, Rahul K and Bai, Yu and Baker, Bowen and Bao, Haiming and others},
  journal={arXiv preprint arXiv:2508.10925},
  year={2025}
}

@article{liu2023g,
  title={G-eval: NLG evaluation using gpt-4 with better human alignment},
  author={Liu, Yang and Iter, Dan and Xu, Yichong and Wang, Shuohang and Xu, Ruochen and Zhu, Chenguang},
  journal={arXiv preprint arXiv:2303.16634},
  year={2023}
}

@inproceedings{huang2023humanity,
  title={On the humanity of conversational ai: Evaluating the psychological portrayal of llms},
  author={Huang, Jen-tse and Wang, Wenxuan and Li, Eric John and Lam, Man Ho and Ren, Shujie and Yuan, Youliang and Jiao, Wenxiang and Tu, Zhaopeng and Lyu, Michael},
  booktitle={The Twelfth International Conference on Learning Representations},
  year={2023}
}

@article{bi2024deepseek,
  title={Deepseek llm: Scaling open-source language models with longtermism},
  author={Bi, Xiao and Chen, Deli and Chen, Guanting and Chen, Shanhuang and Dai, Damai and Deng, Chengqi and Ding, Honghui and Dong, Kai and Du, Qiushi and Fu, Zhe and others},
  journal={arXiv preprint arXiv:2401.02954},
  year={2024}
}

@article{team2024gemma,
  title={Gemma 2: Improving open language models at a practical size},
  author={Team, Gemma and Riviere, Morgane and Pathak, Shreya and Sessa, Pier Giuseppe and Hardin, Cassidy and Bhupatiraju, Surya and Hussenot, L{\'e}onard and Mesnard, Thomas and Shahriari, Bobak and Ram{\'e}, Alexandre and others},
  journal={arXiv preprint arXiv:2408.00118},
  year={2024}
}

@article{ahmed2025qwen,
  title={Qwen 2.5: A comprehensive review of the leading resource-efficient llm with potentioal to surpass all competitors},
  author={Ahmed, Imtiaz and Islam, Sadman and Datta, Partha Protim and Kabir, Imran and Chowdhury, Naseef Ur Rahman and Haque, Ahshanul},
  journal={Authorea Preprints},
  year={2025},
  publisher={Authorea}
}

@inproceedings{giunchi2024dreamcodevr,
  title={Dreamcodevr: Towards democratizing behavior design in virtual reality with speech-driven programming},
  author={Giunchi, Daniele and Numan, Nels and Gatti, Elia and Steed, Anthony},
  booktitle={2024 IEEE Conference Virtual Reality and 3D User Interfaces (VR)},
  pages={579--589},
  year={2024},
  organization={IEEE}
}

@article{brooke1996sus,
  title={SUS-A quick and dirty usability scale},
  author={Brooke, John and others},
  journal={Usability evaluation in industry},
  volume={189},
  number={194},
  pages={4--7},
  year={1996},
  publisher={London, England}
}

@inproceedings{schmidmaier2024perceived,
  title={Perceived Empathy of Technology Scale (PETS): measuring empathy of systems toward the user},
  author={Schmidmaier, Matthias and Rupp, Jonathan and Cvetanova, Darina and Mayer, Sven},
  booktitle={Proceedings of the 2024 CHI Conference on Human Factors in Computing Systems},
  pages={1--18},
  year={2024}
}

@article{bartneck2008godspeed,
  title={Godspeed questionnaire series},
  author={Bartneck, Christoph and Kuli{\'c}, Dana and Croft, Elizabeth and Zoghbi, Susana},
  journal={International journal of social robotics},
  year={2008}
}

@incollection{bartneck2023godspeed,
  title={Godspeed questionnaire series: Translations and usage},
  author={Bartneck, Christoph},
  booktitle={International handbook of behavioral health assessment},
  pages={1--35},
  year={2023},
  publisher={Springer}
}

@article{russell1980circumplex,
  title={A circumplex model of affect.},
  author={Russell, James A},
  journal={Journal of personality and social psychology},
  volume={39},
  number={6},
  pages={1161},
  year={1980},
  publisher={American Psychological Association}
}

@article{zaharieva2024automated,
  title={Automated facial expression measurement in a longitudinal sample of 4-and 8-month-olds: Baby FaceReader 9 and manual coding of affective expressions},
  author={Zaharieva, Martina S and Salvadori, Eliala A and Messinger, Daniel S and Visser, Ingmar and Colonnesi, Cristina},
  journal={Behavior research methods},
  volume={56},
  number={6},
  pages={5709--5731},
  year={2024},
  publisher={Springer}
}

@misc{Li2025EmoVerse,
  author    = {Li, A. and Xu, L. and Ling, C. and Zhang, J. and Wang, P.},
  title     = {EmoVerse: Exploring Multimodal Large Language Models for Sentiment and Emotion Understanding},
  year      = {2025},
  eprint    = {2412.08049},
  archivePrefix = {arXiv},
  primaryClass = {cs.CL}
}

@inproceedings{Liu2024OpenSet,
  author    = {Liu, Y. and Huang, Y. and Liu, S. and Zhan, Y. and Chen, Z. and Chen, Z.},
  title     = {Open-Set Video-based Facial Expression Recognition with Human Expression-Sensitive Prompting},
  booktitle = {Proceedings of ACM Multimedia '24},
  year      = {2024},
  publisher = {ACM}
}

@misc{Han2024Knowledge,
  author    = {Han, B. and Yau, C. and Lei, S. and Gratch, J.},
  title     = {Knowledge-based Emotion Recognition using Large Language Models},
  year      = {2024},
  eprint    = {2408.04123},
  archivePrefix = {arXiv},
  primaryClass = {cs.CL}
}

@article{nielsen1995applying,
  title={Applying discount usability engineering},
  author={Nielsen, Jakob},
  journal={IEEE software},
  volume={12},
  number={1},
  pages={98--100},
  year={1995},
  publisher={IEEE}
}

@inproceedings{sohaib2010integrating,
  title={Integrating usability engineering and agile software development: A literature review},
  author={Sohaib, Osama and Khan, Khalid},
  booktitle={2010 international conference on Computer design and applications},
  volume={2},
  pages={V2--32},
  year={2010},
  organization={IEEE}
}

@Incollection{Cooper2016Empathy,
  author    = {Bridget Cooper},
  title     = {Empathy, emotion, technology, and learning},
  booktitle = {Emotions, Technology, and Learning},
  editor    = {Sharon Y. Tettegah and Michael P. McCreery},
  series    = {Emotions and Technology},
  publisher = Academic,
  year      = {2016},
  pages     = {265--288},
  doi       = {10.1016/B978-0-12-800649-8.00011-0},
  url       = {https://www.sciencedirect.com/science/article/pii/B9780128006498000110}
}

@article{li2023large,
  title={Large language models understand and can be enhanced by emotional stimuli},
  author={Li, Cheng and Wang, Jindong and Zhang, Yixuan and Zhu, Kaijie and Hou, Wenxin and Lian, Jianxun and Luo, Fang and Yang, Qiang and Xie, Xing},
  journal={arXiv preprint arXiv:2307.11760},
  year={2023}
}

@inproceedings{xiao2020if,
  title={If I hear you correctly: Building and evaluating interview chatbots with active listening skills},
  author={Xiao, Ziang and Zhou, Michelle X and Chen, Wenxi and Yang, Huahai and Chi, Changyan},
  booktitle={Proceedings of the 2020 CHI Conference on Human Factors in Computing Systems},
  pages={1--14},
  year={2020}
}

@article{inzlicht2024praise,
  title={In praise of empathic AI},
  author={Inzlicht, Michael and Cameron, C Daryl and D’Cruz, Jason and Bloom, Paul},
  journal={Trends in Cognitive Sciences},
  volume={28},
  number={2},
  pages={89--91},
  year={2024},
  publisher={Elsevier}
}

@article{decety2014complex,
  title={The complex relation between morality and empathy},
  author={Decety, Jean and Cowell, Jason M},
  journal={Trends in cognitive sciences},
  volume={18},
  number={7},
  pages={337--339},
  year={2014},
  publisher={Elsevier}
}

@article{eklund2021toward,
  title={Toward a consensus on the nature of empathy: A review of reviews},
  author={Eklund, Jakob H{\aa}kansson and Meranius, Martina Summer},
  journal={Patient Education and Counseling},
  volume={104},
  number={2},
  pages={300--307},
  year={2021},
  publisher={Elsevier}
}

@article{haakansson2003empathy,
  title={Empathy as an interpersonal phenomenon},
  author={H{\aa}kansson, Jakob and Montgomery, Henry},
  journal={Journal of Social and Personal Relationships},
  volume={20},
  number={3},
  pages={267--284},
  year={2003},
  publisher={Sage Publications}
}

@article{haase1972nonverbal,
  title={Nonverbal components of empathic communication.},
  author={Haase, Richard F and Tepper, Donald T},
  journal={Journal of counseling psychology},
  volume={19},
  number={5},
  pages={417},
  year={1972},
  publisher={American Psychological Association}
}

@article{mehrabian1967decoding,
  title={Decoding of inconsistent communications},
  author={Mehrabian, Albert},
  journal={Journal of Personality and Social Psychology},
  volume={6},
  number={1},
  pages={109--114},
  year={1967},
  doi={10.1037/h0024532}
}

@article{hall1995nonverbal,
  title={Nonverbal behavior in clinician-patient interaction},
  author={Hall, Judith A. and Harrigan, Jinni A. and Rosenthal, Robert},
  journal={Applied and Preventive Psychology},
  volume={4},
  number={1},
  pages={21--37},
  year={1995},
  doi={10.1016/S0962-1849(05)80049-6}
}

@article{bavelas2000listeners,
  title={Listeners as co-narrators},
  author={Bavelas, Janet Beavin and Coates, Linda and Johnson, Trudy},
  journal={Journal of Personality and Social Psychology},
  volume={79},
  number={6},
  pages={941--952},
  year={2000},
  doi={10.1037/0022-3514.79.6.941}
}

@article{chartrand2013antecedents,
  title={The antecedents and consequences of human behavioral mimicry},
  author={Chartrand, Tanya L. and Lakin, Jessica L.},
  journal={Annual Review of Psychology},
  volume={64},
  pages={285--308},
  year={2013},
  doi={10.1146/annurev-psych-113011-143754}
}

@article{tickle1990nature,
  title={The nature of rapport and its nonverbal correlates},
  author={Tickle-Degnen, Linda and Rosenthal, Robert},
  journal={Psychological Inquiry},
  volume={1},
  number={4},
  pages={285--293},
  year={1990},
  doi={10.1207/s15327965pli0104_1}
}

@article{ayers2023comparing,
  title = {Comparing Physician and Artificial Intelligence Chatbot Responses to Patient Questions Posted to a Public Social Media Forum},
  author = {Ayers, John W. and Poliak, Adam and Dredze, Mark and Leas, Eric C. and Zhu, Zechariah and Kelley, Jessica B. and Faix, Dennis J. and Goodman, Aaron M. and Longhurst, Christopher A. and Hogarth, Michael and Smith, Davey M.},
  journal = {JAMA Internal Medicine},
  year = {2023},
  volume = {183},
  number = {6},
  pages = {589--596},
  doi = {10.1001/jamainternmed.2023.1838}
}

@article{abd-alrazaq2020effectiveness,
  title = {Effectiveness and Safety of Using Chatbots to Improve Mental Health: Systematic Review and Meta-Analysis},
  author = {Abd-Alrazaq, Alaa A. and Rababeh, Asma and Alajlani, Mohannad and Bewick, Bridgette M. and Househ, Mowafa},
  journal = {Journal of Medical Internet Research},
  year = {2020},
  volume = {22},
  number = {7},
  pages = {e16021},
  doi = {10.2196/16021}
}

@article{kasneci2023chatgpt,
  title = {ChatGPT for Good? On Opportunities and Challenges of Large Language Models for Education},
  author = {Kasneci, Enkelejda and Sessler, Kathrin and Küchemann, Stefan and Bannert, Maria and Dementieva, Daryna and Fischer, Frank and Gasser, Urs and Groh, Georg and Günnemann, Stephan and Hüllermeier, Eyke and others},
  journal = {Learning and Individual Differences},
  year = {2023},
  volume = {103},
  pages = {102274},
  doi = {10.1016/j.lindif.2023.102274}
}

@article{kuhail2023interacting,
  title = {Interacting with Educational Chatbots: A Systematic Review},
  author = {Kuhail, Mohammad Amin and Alturki, Nazik and Alramlawi, Salwa and Alhejori, Kholood},
  journal = {Education and Information Technologies},
  year = {2023},
  volume = {28},
  number = {1},
  pages = {973--1018},
  doi = {10.1007/s10639-022-11177-3}
}

@incollection{clark1991grounding,
  title     = {Grounding in Communication},
  author    = {Clark, Herbert H. and Brennan, Susan E.},
  booktitle = {Perspectives on Socially Shared Cognition},
  editor    = {Resnick, Lauren B. and Levine, John M. and Teasley, Stephanie D.},
  pages     = {127--149},
  publisher = {American Psychological Association},
  address   = {Washington, DC},
  year      = {1991},
  doi       = {10.1037/10096-006}
}

@misc{Anonymous2025EmpathicPrompting,
  author       = {Anonymous},
  title        = {Empathic Prompting: Non-Verbal Context Integration for Multimodal LLM Conversations},
  year         = {2025},
}

@inproceedings{dongre2024physiology,
  author = {Dongre, Poorvesh},
  title = {Physiology-Driven Empathic Large Language Models (EmLLMs) for Mental Health Support},
  year = {2024},
  isbn = {9798400704824},
  publisher = {Association for Computing Machinery},
  address = {New York, NY, USA},
  url = {https://doi.org/10.1145/3613905.3651132},
  doi = {10.1145/3613905.3651132},
  booktitle = {Extended Abstracts of the 2024 CHI Conference on Human Factors in Computing Systems},
  articleno = {LBW161},
  numpages = {7},
  location = {Honolulu, HI, USA},
  series = {CHI EA '24}
}

@inproceedings{neupane2025wearable,
  author = {Neupane, Sameer and Dongre, Poorvesh and Gracanin, Denis and Kumar, Santosh},
  title = {Wearable Meets {LLM} for Stress Management: A Duoethnographic Study Integrating Wearable-Triggered Stressors and {LLM} Chatbots for Personalized Interventions},
  year = {2025},
  isbn = {9798400257272},
  publisher = {Association for Computing Machinery},
  address = {New York, NY, USA},
  url = {https://doi.org/10.1145/3706599.3720197},
  doi = {10.1145/3706599.3720197},
  booktitle = {Proceedings of the CHI Conference on Human Factors in Computing Systems},
  articleno = {415},
  numpages = {18},
  location = {Yokohama, Japan},
  series = {CHI '25}
}

@article{thabane2010tutorial,
  title={A tutorial on pilot studies: the what, why and how},
  author={Thabane, Lehana and Ma, Jinhui and Chu, Rong and Cheng, Ji and Ismaila, Afisi and Rios, Lorena P and Robson, Reid and Thabane, Marroon and Giangregorio, Lora and Goldsmith, Charles H},
  journal={BMC medical research methodology},
  volume={10},
  number={1},
  pages={1},
  year={2010},
  publisher={Springer}
}

@article{eldridge2016defining,
  title={Defining feasibility and pilot studies in preparation for randomised controlled trials: development of a conceptual framework},
  author={Eldridge, Sandra M and Lancaster, Gillian A and Campbell, Michael J and Thabane, Lehana and Hopewell, Sally and Coleman, Claire L and Bond, Christine M},
  journal={PloS one},
  volume={11},
  number={3},
  pages={e0150205},
  year={2016},
  publisher={Public Library of Science San Francisco, CA USA}
}

@article{paech2023eq,
  title={Eq-bench: An emotional intelligence benchmark for large language models},
  author={Paech, Samuel J},
  journal={arXiv preprint arXiv:2312.06281},
  year={2023}
}

@article{taber2018use,
  title={The use of Cronbach’s alpha when developing and reporting research instruments in science education},
  author={Taber, Keith S},
  journal={Research in science education},
  volume={48},
  number={6},
  pages={1273--1296},
  year={2018},
  publisher={Springer}
}

@inproceedings{sauro2011designing,
  title={When designing usability questionnaires, does it hurt to be positive?},
  author={Sauro, Jeff and Lewis, James R},
  booktitle={Proceedings of the SIGCHI conference on human factors in computing systems},
  pages={2215--2224},
  year={2011}
}

@article{skantze2021turn,
  title={Turn-taking in conversational systems and human-robot interaction: a review},
  author={Skantze, Gabriel},
  journal={Computer Speech \& Language},
  volume={67},
  pages={101178},
  year={2021},
  publisher={Elsevier}
}

@article{bi2025gptossgood,
  title         = {Is GPT-OSS Good? A Comprehensive Evaluation of OpenAI's Latest Open Source Models},
  author        = {Bi, Ziqian and Chen, Keyu and Tseng, Chiung-Yi and Zhang, Danyang and Wang, Tianyang and Luo, Hongying and Chen, Lu and Huang, Junming and Guan, Jibin and Hao, Junfeng and Song, Xinyuan and Song, Junhao},
  year          = {2025},
  eprint        = {2508.12461},
  archivePrefix = {arXiv},
  primaryClass  = {cs.CL},
  doi           = {10.48550/arXiv.2508.12461},
  url           = {https://arxiv.org/abs/2508.12461}
}

@article{li2024llmsasjudges,
  title         = {LLMs-as-Judges: A Comprehensive Survey on LLM-based Evaluation Methods},
  author        = {Li, Haitao and Dong, Qian and Chen, Junjie and Su, Huixue and Zhou, Yujia and Ai, Qingyao and Ye, Ziyi and Liu, Yiqun},
  year          = {2024},
  eprint        = {2412.05579},
  archivePrefix = {arXiv},
  primaryClass  = {cs.CL},
  doi           = {10.48550/arXiv.2412.05579},
  url           = {https://arxiv.org/abs/2412.05579}
}

@article{terzis2013measuring,
  title={Measuring instant emotions based on facial expressions during computer-based assessment},
  author={Terzis, Vasileios and Moridis, Christos N and Economides, Anastasios A},
  journal={Personal and ubiquitous computing},
  volume={17},
  number={1},
  pages={43--52},
  year={2013},
  publisher={Springer}
}

@article{landmann2023can,
  title={I can see how you feel—Methodological considerations and handling of Noldus's FaceReader software for emotion measurement},
  author={Landmann, Elisa},
  journal={Technological Forecasting and Social Change},
  volume={197},
  pages={122889},
  year={2023},
  publisher={Elsevier}
}

@article{wu2022novel,
  title={A novel markovian framework for integrating absolute and relative ordinal emotion information},
  author={Wu, Jingyao and Dang, Ting and Sethu, Vidhyasaharan and Ambikairajah, Eliathamby},
  journal={IEEE Transactions on Affective Computing},
  volume={14},
  number={3},
  pages={2089--2101},
  year={2022},
  publisher={IEEE}
}

@article{han2025judgesverdict,
  title         = {Judge's Verdict: A Comprehensive Analysis of LLM Judge Capability Through Human Agreement},
  author        = {Han, Steve and Titericz Junior, Gilberto and Balough, Tom and Zhou, Wenfei},
  year          = {2025},
  eprint        = {2510.09738},
  archivePrefix = {arXiv},
  primaryClass  = {cs.CL},
  doi           = {10.48550/arXiv.2510.09738},
  url           = {https://arxiv.org/abs/2510.09738}
}

@misc{OpenFaceOpenFace,
title = {OpenFace: a state-of-the-art tool for facial landmark detection, head pose estimation, facial action unit recognition, and eye-gaze estimation},
author = {Baltrušaitis, Tadas and Zadeh, Amir and Lim, Yao Chong and Morency, Louis-Philippe},
howpublished = {GitHub repository},
year = {2018},
url = {https://github.com/TadasBaltrusaitis/OpenFace},
}

@misc{PyFeatPyFeat,
title = {Py-Feat: Python Facial Expression Analysis Toolbox},
author = {Py-Feat authors},
howpublished = {Web resource},
year = {2020},
url = {https://py-feat.org/pages/intro.html},
}

@misc{MediaPipeMediaPipe,
title = {MediaPipe: Cross-platform ML solutions for live and streaming media},
author = {Google AI Edge},
howpublished = {GitHub repository and documentation},
year = {2025},
url = {https://github.com/google-ai-edge/mediapipe},
}

@article{Kurdi2017OASIS,
  author  = {Kurdi, Benedek and Lozano, Shayn and Banaji, Mahzarin R.},
  title   = {Introducing the Open Affective Standardized Image Set (OASIS)},
  journal = {Behavior Research Methods},
  year    = {2017},
  volume  = {49},
  number  = {2},
  pages   = {457--470},
  doi     = {10.3758/s13428-016-0715-3}
}






\end{document}